\documentclass[a4paper,12pt]{article}

\usepackage{amsmath,amssymb}

\def\S3{{\sqrt{3}}}
\def \hring {{\hat{\cal R}}}
\def \ring {{\cal R}}

\def \sign {\, \mbox{sign}}

\def \spc {\, \mbox{Span}_{\bbbc}}
\def\diag {\mbox{diag}}

\def\bbbn{{\mathbb N}}
\def\bbbc{{\mathbb C}}
\def\bbbz{{\mathbb Z}}

\def\bbbq{{\mathbb Q}}

\def\bu{{\bf u}}
\def\bF{{\bf F}}
\def\bG{{\bf G}}
\def\bH{{\bf H}}
\def\zp{{{\mathbb Z}_{\ge 0}}}
\def\zpi{{{\mathbb Z}_{\ge 0}^{\infty *}}}
\def\zpy{{{\mathbb Z}_{\ge 0}^{\infty}}}

\def\hu{{\hat{u}}}
\def\hv{{\hat{v}}}

\def\cL{{\cal L}}

\def\cR{{\cal R}}

\def\cS{{\cal S}}
\def\cW{{\cal W}}
\def\s5{\sqrt{5}}

\def\tr{\mbox{tr}}

\newtheorem{Def}{Definition}
\newtheorem{The}{Theorem}
\newtheorem{Pro}{Proposition}

\newtheorem{Cor}{Corollary}

\begin{document}
\author{Alexander.V. Mikhailov\\School of Mathematics, University of Leeds, UK
\and Vladimir S.~Novikov\\School of Mathematics, Loughborough University, UK \and
Jing Ping Wang\\IMSAS, University of Kent, UK\and\\
}

\title{Symbolic representation and classification of integrable systems}

\maketitle

\section{Introduction}
Among  nonlinear partial differential equations (PDEs) there is
an exceptional class of so-called {\em integrable} equations. Integrable equations
can be studied with the same completeness as linear PDEs, at least in principle. They
possess a rich set of exact solutions and many hidden properties such as infinite
hierarchies of symmetries, conservation laws, etc.

There are two known classes of integrable equations. One class is linearisable
equations, i.e. equations related to linear ones by differential substitutions. For
example, the famous Burgers equation
\[ u_t=u_{xx}+2uu_x\]
can be linearised by a differential substitution $u=\phi_x/\phi$ (the Cole-Hopf
transformation). In the new variable $\phi$ it takes the form of a linear heat
equation:
\[ \phi_t=\phi_{xx} \, .\]
Another class is equations solvable by the inverse scattering transform method
(such as the Korteweg de--Vries equation $u_t=u_{xxx}+6uu_x$). There is massive
literature on integrable equations, their solutions and properties (see for example
monographs \cite{zmnp, AS, ne85, mr93g:35108, mr94g:58260} and references therein).

In the {\em symmetry approach} the existence of higher symmetries of a PDE, or more
precisely, of an infinite hierarchy of higher symmetries is regarded as the
definition of its integrability. Existence of a finite number of symmetries of a
partial differential equation may not secure its integrability
\cite{Ba91,mr99i:35005,mr2003j:37109,mnw05}.

The symmetry approach based on a concept of formal recursion operator has been
formulated and developed in works of Shabat and co-authors (see for example review
papers \cite{mr86i:58070,mr89e:58062,mr93b:58070,mr95j:35010}). It has been shown
that the existence of an infinite hierarchy of symmetries or local conservation laws
or a possibility to linearise a certain equation imply the existence of a formal
recursion operator. Formal recursion operator is carrying information about
integrability and is  not sensitive to lacunas in the infinite hierarchy of
symmetries or conservation laws. The conditions of its existence give integrability
conditions for the equation which can be formulated in a form of an infinite sequence
of canonical densities and encode many hidden properties of the equation. If a
density is non-trivial (i.e. is not a total derivative) it provides a density of a
local conservation law of the equation. For linearisable equations all densities
except a finite number are trivial. The sequence of canonical densities is invariant
with respect to invertible (and almost invertible \cite{mr89g:58092})
transformations.

It was the
first approach enabling  to give a complete description of integrable evolutionary
equations of the form
\begin{eqnarray*}
 u_t&=&f(u_{xx},u_x,u,x,t)\, ,\\
u_t&=&\partial^n_xu+f(\partial^{n-1}_xu,\ldots,\, u_x,\, u)\,,\quad n=3,4,5\, ,
\end{eqnarray*}
where $f$ is a smooth function of its arguments
\cite{svin-BUR,mr84j:35152,mr87d:58071}. Integrable differential-difference equation
of the form
\[
 u_{n,\, t}=f(u_{n-1},u_n,u_{n+1})\,, \qquad n\in\bbbz
\]
have been classified in \cite{yamilov,mr98i:58206}.
It can also be applied to systems of
equations. In particularly a complete classification of systems of two equations of
the form
\begin{equation}\label{sys2ms}
\bu_t=A(\bu )\bu_{xx}+{\bf F}(\bu ,\bu_x),\quad \det (A(\bu ))\ne 0,\quad \bu=(u,v)^T
\end{equation}
possessing infinite hierarchies of local conservation laws has been given in
\cite{mr87h:35313,mr87h:35312,mr89g:58092}. Existence of higher conservation laws
immediately implies that the trace of $A(\bu)$ vanishes. System (\ref{sys2ms}) can
still possess an infinite hierarchy of symmetries even if the trace of $A(\bu)$ is
not equal to zero. In the latter case the system is linearisable by a Cole-Hopf type
differential substitution. In the case of  $A(\bu )$ being a unit matrix integrable
systems of the type (\ref{sys2ms}) have been listed in \cite{mr90a:35212}. A complete
classification of polynomial homogeneous integrable system of the form (\ref{sys2ms})
in the case when matrix $A(\bu )$ is constant and has two distinct eigenvalues is
given in \cite{MR2070382}.

In this approach it is not assumed that equations are polynomial or rational
functions of independent variables and their derivatives. A disadvantage is that for
every fixed order of the equation the integrability conditions have to be derived
from scratch and thus it is difficult to draw a global picture, i.e., in all orders.
Also, this approach is heavily based on the concept of locality which makes it
difficult to apply to integro-differential, non-evolutionary and multi-dimensional
systems.

In this article we would like to give a brief account of recent development of the
symmetry approach. The progress has been achieved mainly due to a symbolic
representation of the ring of differential polynomials which enable us to use
powerful results from algebraic geometry and number theory. Symbolic representation
(an abbreviated form of the Fourier transformation) has been originally applied to
the theory of integrable equations by Gel'fand and Dikii \cite{mr58:22746}.
Symmetry approach in symbolic representation has been formulated
and developed to the problem of the global classification of integrable evolutionary
equations in \cite{wang98,mr99g:35058,mr2001h:37147}. In symbolic representation the
existence of infinite hierarchy of symmetries is linked with factorisation properties
of an infinite sequence of multi-variable polynomials. Symbolic
representation is a suitable tool to study integrability of noncommutative
\cite{mr1781148}, non-evolutionary \cite{mr1908645,mnw07,nw07}, non-local
(integro--differential) \cite{mn2}, multi--component
\cite{MR2070382,mr1829636,kamp02a} and multi--dimensional equations \cite{wang21}. It
is convenient for testing integrability of a given system, provides useful
information on the structure of the symmetry hierarchy and suitable for global
classification of integrable equations. In this framework it is natural to define
approximate symmetries and  approximate integrability. Study and classification of
approximately integrable equations is a new and unexplored area of research with a
considerable potential for applications. Symmetry approach in symbolic representation
has certain drawbacks due to a restriction to the ring of differential polynomials,
which can be amended in some cases by suitable extensions of the ring.

In literature one can find other attempts to describe integrable systems based on
properties of solutions, such as the Painlev\'e property in the analytical theory
(see for example \cite{AS, ne85, mr93g:35108}), existence of three soliton solutions
for bi-linear (Hirota) representation (see the article by Hietarinta in this book),
elasticity of soliton collisions in numerical experiments, etc. Existence of one
nontrivial symmetry of an prior fixed order can be used for isolation of integrable
equations \cite{mr81i:35144,mr2002j:35275}. Every method has certain advantages and
disadvantages. The reader can judge the power of each approach by the results
obtained, their completeness and generality.

Our article is organised as following. In the next Section \ref{sect2} we give basic definitions and notations. We define symmetries, approximate symmetries and formal recursion operator. Then we introduce symbolic representation of the ring of differential polynomials. A generalisation to several dependent variables is given in the last part of this section. In Section \ref{sec3}, using approximate symmetries in symbolic representation, we study the structure of the Lie algebra of symmetries. In symbolic representation the existence of approximate symmetries can be reformulated in terms of factorisation properties of polynomials. Existence of one nontrivial symmetry enables us to constructively extend any approximate symmetry of degree 3 to any degree (Theorem \ref{MainR}). In symbolic representation  conditions for the existence of formal recursion operator lead to a simple test for integrability. Here a state of art result is a global classification of integrable homogeneous evolutionary equations (Section \ref{sec_glob}). The result of the
classification can be accounted as following: {\sl Integrable equations are
symmetries (members of infinite hierarchies) of nonlinear PDEs of orders $2,\ 3$ or
$5$. Thus it is sufficient to classify integrable equations of order $2,\ 3$ or $5$.
There is only a finite number of such equations (namely $10$) and the corresponding
hierarchies of symmetries} (Theorem \ref{globclass}). In Section \ref{sec_nonlocal} we apply our method to non-local (Benjamin-Ono and Camassa-Holm type) equations. It requires a non-local extension of the ring of differential polynomials and  symbolic representation proved to be a suitable language to tackle the problem. In Section \ref{sec_bouss} we present results on classification of Boussinesq type equations. In the case of even order equations we use conditions following from the existence of formal recursion operator. Together with classification results for orders 4 and 6 we  present three new integrable equations of order 10. A global classification of all integrable odd order Boussinesq type equation is given in section \ref{oddorderbousseq}. A generalisation of the symbolic approach to $2+1$--dimensional equations enables us (Section \ref{2+1}) to study the structure of symmetries of Lax integrable equations and  to prove the conjecture on the structure of non-local terms. Finally we discuss a progress in the problem of classification of systems of integrable equations. In particularly we have found two integrable systems of order 5 which we believe are new.

\section{Symmetries  and formal recursion operators in symbolic representation}\label{sect2}

\subsection{Differential polynomials}\label{diffpoly1}
We shall adopt the following notations: $u_n$ denotes $n$-th derivative $\partial^n_x
u$ of the dependent variable $u$. In particular, $u_0$ denote the function $u$ itself
(often we shall omit the zero index  of $u_0$ and simply write $u$).

A $u$-monomial is a finite product of the form
\begin{equation*}\label{umon}
 u_0^{\alpha_0} u_1^{\alpha_1}\cdots
u_k^{\alpha_k},\end{equation*} where all exponents
$\alpha_0,\ldots,\alpha_k$ are non-negative integers
($\alpha_s\in\zp$), and the total degree is
$|\alpha|=\alpha_0+\alpha_1+\cdots \alpha_k >0$.

A finite sequence  $\alpha=(\alpha_0,\ldots ,\alpha_k)$ can be seen as an element of
a semi-group $\zpi$ of infinite sequences of non-negative integers, such that only a
finite number of entries in a sequence are non-zero and there is at least one nonzero
entry. There is an obvious bijection between the set of all $u$-monomials and $\zpi$.
We can simplify the notations as follows:
\[ u^{\alpha}=u_0^{\alpha_1} u_1^{\alpha_2}\cdots u_k^{\alpha_k}.\]

A {\em differential polynomial} $f$ in variables $u_0,u_1,\ldots$ with coefficients
in $\bbbc$ is a finite linear combination of $u$-monomials, i.e.,
\[ f=\sum_{\alpha\in A}a_{\alpha}u^\alpha \, , \qquad a_\alpha \in \bbbc,\]
where the sum is over a finite set $A=\{\alpha\, |\, \alpha\in\zpi \}$. The set of
all such differential polynomials is denoted $\cR$. It is a
 ring with usual addition and multiplication of polynomials.
Moreover, it is a differential ring  and $\bbbc\not\in\ring$. The linear operator
\begin{equation}\label{Dx} D_x =\sum_{k\ge 0}\left(u_{k+1}\frac{\partial}
{\partial u_k}\right) \end{equation} is a derivation of $\cR$ corresponding to the
total $x$-derivative. Operator $D_x$ acting on an element  $f\in\cR$ results in a
finite sum depending on the choice of $f$. Therefore, we do not indicate the upper
limit for the summation in the definition of $D_x$ and all other operators defined in
this article.

It is easy to verify that monomials $u^{\alpha}$ are eigenvectors of the following
commuting linear operators
\begin{equation}
\label{DuX} D_u=\sum_{k\ge 0}u_{k}\frac{\partial}{\partial u_k}, \qquad
X_u=\sum_{k\ge 1}\left(ku_{k}\frac{\partial}{\partial u_k}\right)
\end{equation}
with
 \( D_u(u^{\alpha})=|\alpha|u^{\alpha},\) and \( X_u(u^{\alpha})=d_{\alpha} u^{\alpha},
\) where $ d_{\alpha}=\sum_{k\ge 1} (k\alpha_k)\, .$

Thus the ring $\cR$ is graded and is a direct sum of eigenspaces
\[ \cR=\bigoplus_{n\in\bbbn}\cR^{n}=\bigoplus_{n,p\in\bbbn}\cR_{p-1}^{n},\]
\[ \cR^{n}=\{f\in\cR\, |\, D_u(f)=nf\}\, ,\quad  \cR_p^{n}=\{f\in\cR^n\, |\,
X_u(f)=pf\}. \] If $f\in \cR_p^{n},\ g\in\cR_q^{m}$, then $f\cdot g\in
\cR_{p+q}^{n+m}$. Simply speaking, $\cR_p^{n}$ is a linear subspace of homogeneous
differential polynomials such that each monomial has: (i).  the number of $u$ and its
derivatives being $n$; (ii).  the total number of derivatives being $p$. For example,
\[ u_1^2  u_7+2 u_2 u_3  u_4-u_0^2 u_9\in \cR^{3}_9\subset\cR^3.\]

In some applications it is convenient to introduce weighted homogeneous polynomials.
Let us assume that dependent variable $u$ has a weight $\lambda$ which is a fixed
rational number. We define a linear differential operator
\[ W_{\lambda}=\lambda D_u+X_u,\quad W_{\lambda}:\cR\to\cR .\]
Differential monomials are eigenvectors of $W_{\lambda}$  and the spectrum of
$W_{\lambda}$ is a set $S_{\lambda}=\{n \lambda+m-1\, |\, n,m\in\bbbn\}$. We can
decompose $\cR$ in a direct sum of eigenspaces
\[ \cR=\bigoplus_{\mu\in S_{\lambda}}\cW_\mu\, ,\qquad \cW_\mu=\{f\in\cR\, |\,
W_{\lambda} (f)=\mu f\}\, .  \] Elements of $\cW_\mu$ are called
$\lambda$-homogeneous differential polynomials of weight $\mu$.  For example,
$u_3+6uu_1$ is a $2$-homogeneous differential polynomial of weight $5$. We have
$\cW_\mu \cW_\nu\subset \cW_{\mu+\nu}$. Moreover, if $\lambda>0$ then subspaces
$\cW_\mu$ are finite dimensional.

It is useful to define the little ``${ oh}$'' order symbol.

\begin{Def} Let $f\in\cR$. We say that $f={ o}(\cR^n)$ if
 \( f\in \bigoplus_{k>n}\cR^{k}. \)
\end{Def}

For example, $f={ o}(\cR^3)$ means that the differential polynomial $f$ does not have
linear, quadratic and cubic terms in $u$ and its derivatives.

For any two elements $f,g\in\cR$ we define a Lie bracket
\begin{equation}\label{liebracket}
[f,\ g]=f_*(g)-g_*(f) \, ,
\end{equation}
where the {\em Fr\'echet derivative} for any element $h\in \ring$ is defined as a
linear differential operator of the form
\begin{equation}
h_*=\sum_{k\ge 0}\frac{\partial h}{\partial u_k}D_x^k \, . \label{frechet}
\end{equation}
We say that an element $h\in \ring$ has {\em order} $n$ if the corresponding
differential operator $h_*$ is of order $n$.

Thus $\cR$, treated as a linear space over $\bbbc$, together with the Lie bracket
(\ref{liebracket}) is an infinite dimensional Lie algebra of differential polynomials
over $\bbbc$. The bilinearity and skew-symmetry of the bracket (\ref{liebracket}) are
obvious. The Jacobi identity can be easily verified.

The grading of $\cR$ induces the grading of the Lie algebra of differential
polynomials since we have
\begin{equation}\label{gradLie}
 [\cR^n_p,\cR^m_q]\subset \cR^{n+m-1}_{p+q} \, .
\end{equation}

\subsection{Symmetries, approximate symmetries and formal recursion operator}\label{sec2.2}

In this section, for the sake of simplicity, we give definitions suitable for
evolutionary equations
\begin{equation}\label{eveq1}
  u_t=F,\qquad F\in \cR \, .
 \end{equation}
These definitions will be later extended to non-evolutionary equations and the
multi-component systems of evolutionary equations.

Evolutionary equation (\ref{eveq1}) defines a derivation $D_F:\cR\mapsto \cR$:
\[
 D_F(a)=a_*(F).
\]
In this notation the derivative $D_x=D_{u_1}$ and it is in agreement with
(\ref{DuX}). Sometimes, for simplification of notations, we will denote  $D_F$ as
$D_t$.

\begin{Def}\label{def_sym}
A differential polynomial $G\in\cR$ is said to be a {\sl symmetry} ({\sl a generator
of an infinitesimal symmetry}) for an evolutionary partial differential equation
(\ref{eveq1})
 if the Lie bracket of $F$ and $G$ vanishes, i.e.,
\( [F,\ G]=0 \, .\)
\end{Def}

If $G$ is a symmetry, then evolutionary equation \( u_\tau =G \) is compatible with
(\ref{eveq1}). There are many other equivalent definitions of symmetry (see for
example \cite {mr86i:58070,mr93b:58070}). Elements of $\cR$ do not depend on $x,t$
explicitly. Thus our definition does not include space and time dependent symmetries
such as dilatation and Galilean symmetries. In this article, when we are talking
about symmetries we do mean space and time independent symmetries.

Symmetries form a subalgebra $C_\cR (F)=\{G\in\cR \, |\, [F,\ G]=0\} $ which is the
centraliser of $F$ (it immediately follows from the Jacobi identity). Since $F\in\cR$
does not depend on $x,t$ explicitly, then any equation (\ref{eveq1}) possesses {\em
trivial} symmetries $u_1,F\in C_\cR (F)$ corresponding to translations in space and
time.

\begin{Def}\label{Def3}
Equation (\ref{eveq1}) is said to be {\sl integrable} if its algebra of symmetries
$C_\cR (F)$ is infinite dimensional.
\end{Def}

For nonlinear equations of the form
\begin{eqnarray}\label{ev1}
 u_t=u_n+f(u_{n-1},\ldots ,u) \qquad n\ge 2,
\end{eqnarray}
it is easy to show that the algebra of
symmetries $C_\cR (u_t)$ is commutative. Moreover,
the symmetries $G\in\cR$  must have a linear term (see section \ref{sec24}).

Having in mind the gradation (\ref{gradLie}) of the Lie algebra of differential
polynomials we represent the right hand side of equation (\ref{eveq1}) and
infinitesimal generators of symmetries in the form
\[ F=f_1+f_2+\cdots \,,\quad G=g_1+g_2+\cdots \,,\qquad f_k,g_k\in\cR^k \]
and study them in the sequence of terms, i.e., linear, quadratic, cubic, etc.

\begin{Def} A differential polynomial $G\in\cR$ is said to be an
{\sl approximate symmetry  of degree $p$} for evolutionary partial differential
equation (\ref{eveq1}) if \( [F,\ G]=o(\cR^p). \)
\end{Def}

Equation (\ref{ev1}) possesses infinitely many approximate symmetries of degree 1. An
equation may possess approximate symmetries of degree 2, but fail to possess
approximate symmetries of degree 3. An integrable equation possesses infinitely many
approximate symmetries of any degree. In the next section, using symbolic
representation we formulate the necessary and sufficient conditions for the existence
of approximate symmetries of arbitrary degree.

For example, equation
\begin{equation}\label{contr}
 u_t=u_5+5uu_1 \, ,
\end{equation}
has approximate symmetry of degree $2$ with a generator
\[ G=u_7+7 u u_3 +14 u_1u_2 \, .\]
Indeed,
\[ [G,u_5+5uu_1 ]=210 u_1^2 u_2+105 u u_2^2+105 uu_1u_3
=o(\cR^{2}). \]
Moreover, equation (\ref{contr}) has infinitely many approximate symmetries of degree
$2$ (this fact will become obvious in the next section), but fails to have
approximate symmetries of degree $3$ and thus it is not integrable.

It follows from the Jacobi identity and (\ref{gradLie}) that approximate symmetries
of degree $n$ form a subalgebra of $\cR$ which we denote $C^n_\cR (F)$. Obviously
\begin{equation*}\label{asubal}
\cR=C^1_\cR(F)\supset C^2_\cR(F)\supset C^3_\cR(F)\supset\cdots\supset
 C^\infty_\cR(F)=C_\cR(F)\, .
\end{equation*}

Formal pseudo-differential series, which for simplicity we shall
call formal series, are defined as
\begin{equation}\label{A}
 A=a_{m}D_x^m+a_{m-1}D_x^{m-1}+\cdots + a_0+a_{-1}D_x^{-1}+\cdots\,
\quad a_k\in \ring\, .
\end{equation}
The product of two formal series is defined by
\begin{equation}\label{aDbD}
 aD_x^k\circ bD_x^m =a(bD_x^{m+k}+C_k^1 b_1D_x^{k+m-1}+C_k^2 b_{2}D_x^{k+m-2}+
\cdots ) ,
\end{equation}
where $b_{j}=D_x^j(b)$, $k,m\in \mathbb Z$ and the binomial coefficients are defined
as
\[
C^j_n=\frac{n(n-1)(n-2)\cdots(n-j+1)}{j!}\,  .
\]
This product is associative.

\begin{Def}\label{defrecop1} A formal series
\begin{equation}\label{Lm}
\Lambda=l_{m}D_x^m+\cdots +
 l_0+l_{-1} D_x^{-1} + \cdots\,  , \quad l_k\in \ring
\end{equation}
is called a formal recursion operator for equation (\ref{eveq1}) if
\begin{equation}\label{Leq}
D_F(\Lambda)=F_*\circ \Lambda-\Lambda\circ F_*\,  .
\end{equation}
\end{Def}
In literature a formal recursion operator is also called a formal
symmetry of equation (\ref{eveq1}).

The central result of the Symmetry Approach can be represented by the following
Theorem, which we attribute to Shabat:

\begin{The} If equation (\ref{eveq1}) has an infinite hierarchy of
symmetries of arbitrary high order, then  a formal recursion operator exists and its
coefficients can be found recursively.
\end{The}

The Theorem states that for integrable equations, i.e. equations possessing an
infinite hierarchy of higher symmetries, one can solve equation (\ref{Leq}) and
determine recursively the coefficients $l_m,l_{m-1}, \ldots$ of  $\Lambda$ such that
all these coefficients will belong to the ring $\ring$. The solvability conditions of
equation (\ref{Leq}) can be formulated in an elegant form of a canonical sequence of
local conservation laws of equation (\ref{eveq1}). They provide powerful necessary
conditions of integrability. These conditions can be used for testing for
integrability  for a given equation or even for a complete description of integrable
equations of a particular order. A detailed description of the Symmetry Approach
including the proof of the above theorem and classification results for integrable
PDEs and systems of PDEs based on the concept of formal recursion operator can be
found in review papers \cite{mr86i:58070,mr89e:58062,mr93b:58070,mr95j:35010}.

\subsection{Symbolic representation}\label{symbol}
Symbolic representation transforms the problems in differential algebra into the ones
in algebra of symmetric polynomials. This enables us to use powerful results from
Diophantine equations, algebraic geometry and commutative algebra. Symbolic
representation is widely used in theory of pseudo-differential operators. To
integrable systems it was first applied by Gel'fand and Dickey \cite{mr58:22746} and
further developed in works of Beukers, Sanders and Wang \cite{mr99i:35005, wang98,
mr99g:35058}. Symbolic representation can be viewed as a simplified notation for a
Fourier transform \cite{mr1908645}.

In order to define the symbolic representation $\hat{\cR}=\oplus\hat{\cR}^n$ of the
ring (and Lie algebra) of differential polynomials $\cR=\oplus\cR^n$, we first define
an isomorphism of the linear spaces $\cR^n$ and $\hat{\cR}^n$ and then extend it to
isomorphisms of the differential ring and Lie algebra equipping $\hat{\cR}$ with the
multiplication, derivation and Lie bracket.

{\em Symbolic transform} defines a linear isomorphism between the space \(\cR^n\) of
differential polynomials of degree \(n\) and the space \(\bbbc[\xi _1,\ldots
,\xi_{n}]^{\cS^{\xi}_n}\) of algebraic symmetric polynomials in \(n\) variables,
where $\cS^{\xi}_n$ is a permutation group of $n$ variables $\xi_1,\dots ,\xi_n$.
Elements of $\hring^n$ are denoted by $\hu^n a(\xi_1,\ldots,\xi_n)$, where
$a(\xi_1,\ldots,\xi_n)\in \bbbc[\xi _1,\ldots ,\xi_{n}]^{\cS^{\xi}_n}$. The
isomorphism of linear spaces $\cR^n$ and $\hring^n$ is uniquely defined by its action
on monomials.

\begin{Def}\label{repmon}
The {\em symbolic form} of a differential monomial is defined as
\begin{equation*}
u_{i_1}u_{i_2}\cdots u_{i_n}\in\cR^n \quad \longmapsto \quad \hu^{n}
\langle\xi_1^{i_1} \xi_{2}^{i_2} \cdots \xi_{n}^{i_n}\rangle_{\cS^{\xi}_n}\in\hring^n
\end{equation*}
where $\langle\cdot \rangle_{\cS^\xi_n}$ denotes the average over the group
${\cS^\xi_n}$ of permutation of $n$ elements $\xi_1,\ldots ,\xi_n$:
\[  \langle a(\xi_1, \cdots ,\xi_{n})\rangle_{\cS^{\xi}_n}= \frac{1}{n!}
\sum_{\sigma\in \cS^{\xi}_n}
a(\xi_{\sigma (1)}, \cdots , \xi_{\sigma(k)}).
\]
\end{Def}
For example, \begin{eqnarray*} &&u_k\longmapsto \hu\xi_1^k,\  u^n\longmapsto \hu^n,\
uu_1\longmapsto\frac{\hu^2}{2}(\xi_1+\xi_2),\\&&
uu_1^2\longmapsto\frac{\hu^3}{3}(\xi_2 \xi_3+\xi_1 \xi_2+\xi_1 \xi_3),\
u_k^3\longmapsto\hu^3 \xi_1^k \xi_2^k \xi_3^k.\end{eqnarray*}

It is easy to see that in $\hring$ the linear operators $D_u$ and $X_u$, cf.
(\ref{DuX}) are represented by
\[ \hat{D}_u=\hu\frac{\partial}{\partial \hu}\, ,
\qquad \hat{X_u}=\sum_{i=1}\xi_i\frac{\partial}{\partial \xi_i}\,.\] With this
isomorphism the linear spaces $\hring^n_p$ corresponding to $\cR^n_p$ have the
property that the coefficient functions $a(\xi_1,\ldots,\xi_n)$ of symbols $$\hu^n
a(\xi_1,\ldots,\xi_n)\in\hring^n_p$$ are homogeneous symmetric polynomials of degree
$p$.

One of the advantages of the symbolic representation is that the action of the
operator $D_x$, cf. (\ref{Dx}) is very simple. Indeed, let $f\in\cR^n$ and
$f\longmapsto \hu^n a(\xi_1,\ldots,\xi_n)$ then
\[ D_x (f)\quad \longmapsto \quad \hu^n a(\xi_1,\ldots,\xi_n)(\xi_1+\cdots +\xi_n)\, ,\]
and thus $D_x^k (f)\ \longmapsto \ \hu^n a(\xi_1,\ldots,\xi_n)(\xi_1+\cdots
+\xi_n)^k$.

Let $f\in\cR^n,\ f\mapsto \hu^n a(\xi_1,\ldots,\xi_n)$ and $g\in\cR^m,\ g\mapsto
\hu^m b(\xi_1,\ldots,\xi_m)$, then

(i). The product $f\cdot g$ has the following symbolic representation:
\begin{eqnarray*}
f\cdot g \quad &\longmapsto& \quad \hu^n a(\xi_1,\ldots,\xi_n)\circ \hu^m
b(\xi_1,\ldots,\xi_m)\\&=&\hu^{n+m}\langle a(\xi_1,\ldots,\xi_n)
b(\xi_{n+1},\ldots,\xi_{n+m})\rangle_{\cS^{\xi}_{n+m}}.
\end{eqnarray*}
This defines the corresponding multiplication $\circ$ in $\hring$. Representation of
differential monomials (Definition \ref{repmon}) can be deduced from $u_k\longmapsto
\hu \xi_1^k$ and this multiplication rule.

(ii). The Lie bracket $[f,g]$, cf. (\ref{liebracket}) is represented by
\begin{eqnarray}\label{slie}
&&[f,\ g]\quad \longmapsto \quad \hu^{n+m-1}\\
&&\left\langle  n a(\xi_1,\ldots,\xi_{n-1},\xi_n+\cdots
+\xi_{n+m-1})b(\xi_{n},\ldots,\xi_{n+m-1})-
\right.\nonumber \\
&&\left.  m b(\xi_1,\ldots,\xi_{m-1},\xi_m+\cdots +\xi_{n+m-1})
a(\xi_{m},\ldots,\xi_{n+m-1}) \right\rangle _{\cS^{\xi}_{n+m-1}}\nonumber
\end{eqnarray}
For example, if $f\in\cR^1,\ f\mapsto \hu\omega(\xi_1)$ and $g\in\cR^n,\ g\mapsto
\hu^n a(\xi_1,\ldots,\xi_n)$, then
\begin{equation*}\label{lincomm}
[f,g] \longmapsto (\omega(\xi_1+\cdots +\xi_n)-\omega(\xi_1)-\cdots -\omega(\xi_n))\
\hu^n a(\xi_1,\ldots,\xi_n).
\end{equation*}
In particularly, for $f=u_1$ we have $\omega(\xi_1)=\xi_1$ and $[u_1,g]=0$ for any
$g\in\cR$. Thus $u_1$ is a symmetry for any evolutionary equation $u_t=g$.

Symbolic representation of differential operators (such as the Fr\'echet derivative
(\ref{frechet}) and formal series (\ref{A})) is motivated by the theory of linear
pseudo-differential operators in Fourier representation.  To operator $D_x$
(\ref{Dx}) we shall assign a special symbol $\eta$ and the following rules of action
on symbols:
\[ \eta( \hu^n a(\xi_1,\ldots , \xi_n))=\hu^n a(\xi_1,\ldots , \xi_n)\sum_{j=1}^{n}\xi_j \]
and the composition rule
\[
\eta\circ \hu^n a(\xi_1,\ldots , \xi_n)=\hu^n a(\xi_1,\ldots ,
\xi_n)(\sum_{j=1}^{n}\xi_j+\eta )\, .
\]
The latter corresponds to the Leibnitz rule $D_x\circ f=D_x(f)+f D_x$. Now it can be
shown that the composition rule (\ref{aDbD}) can be represented as following. Let we
have two operators $fD_x^q$ and $gD_x^s$ such that  $f$ and $g$ have symbols $\hu^i
a(\xi_1,\ldots ,\xi_i)$ and $\hu^j b(\xi_1,\ldots ,\xi_j)$ respectively. Then
$$fD_x^q\longmapsto \hu^i a(\xi_1,\ldots ,\xi_i)\eta^q,\qquad
gD_x^s\longmapsto \hu^j b(\xi_1,\ldots,\xi_j)\eta^s $$ and
\begin{equation}\label{fDgD}
fD_x^q \circ gD_x^s\longmapsto \hu^{i+j}\langle a(\xi_1,\ldots ,\xi_i)
(\eta+\sum_{m=i+1}^{i+j}\xi_m)^q b(\xi_{i+1},\ldots ,\xi_{i+j})\eta^s \rangle
_{\cS^{\xi}_{i+j}}\, .
\end{equation}
Here the symmetrisation is taken over the group of permutation of all $i+j$ arguments
$\xi_1,\ldots \xi_{i+j}$, the symbol $\eta$ is not included in this set.  In
particularly, it follows from (\ref{fDgD}) that $D_x^q\circ D_x^s\mapsto \eta^{q+s}$.
The composition rule (\ref{fDgD}) is valid for both positive and negative exponents.
In the case of positive exponents it is a polynomial in $\eta$ and the result is a
Fourier image of a differential operator. In the case of negative exponents one can
expand the result on $\eta$ at $\eta\to \infty$ in order to identify it with
(\ref{aDbD}). In the symbolic representation instead of formal series (\ref{A}) it is
natural to consider a more general object, namely formal series of the form
\begin{equation*}\label{B}
B=b(\eta)+\hu b_1(\xi_1,\eta)+\hu^2 b_2(\xi_1,\xi_2,\eta)+\hu^3
b_3(\xi_1,\xi_2,\xi_3,\eta)+\cdots\, ,
\end{equation*}
where coefficients $b(\eta)\not=0, b_k(\xi_1,\ldots ,\xi_k,\eta)$ are rational
functions of its arguments (with certain restrictions which will be discussed in the
next section).

The symbolic representation of the Fr\'echet derivative of the element $f\longmapsto
\hu^n a(\xi_1,\ldots ,\xi_n)$ is
\[ f_*\longmapsto n \hu^{n-1}a(\xi_1,\ldots ,\xi_{n-1},\eta )\, .\]
For example, let $F=u_3+6uu_1$, then $F\mapsto \hu \xi_1^3+3 \hu^2 (\xi_1+\xi_2)$ and
\[ F_*\mapsto \eta^3+6 \hu (\xi_1+\eta )\, .\]
It is interesting to notice that the symbol of the Fr\'echet derivative is always
symmetric with respect to all permutations of arguments, including the argument
$\eta$. Moreover, the following obvious but useful Proposition holds \cite{mr1908645}:

\begin{Pro} A differential operator is a Fr\'echet derivative of an element
of $\ring$ if and only if its symbol is invariant with respect to
all permutations of its argument, including the argument $\eta$.
\end{Pro}

The symbolic representation has been extended and proved to be
very useful in the case of noncommutative differential rings
\cite{mr1781148}. In the next sections symbolic representation will be
extended to the cases of many
dependent variables, suitable for study of system of
equations and further to the cases of
non-local and multidimensional equations.

\subsection{Generalisation to several dependent variables}\label{severalvar}

The definitions and most of the statements formulated in the previous sections
\ref{diffpoly1}--\ref{sec24} can be easily extended to several dependents, i.e., to
systems of equations. In this section we will give a brief account of definitions and
some results concerning two dependent variables. A generalisation for $N$ dependent
variables is straightforward. For details see \cite{mr89e:58062, mr93b:58070,
mr95j:35010, mnw07, nw07, mr1829636}.

Similarly to $u$--monomials (Section \ref{diffpoly1}), we define $v$-monomials
$$v^\beta =v_0^{\beta_0} v_1^{\beta_1}\cdots v_s^{\beta_s}, \quad \beta\in\zpy.$$
A differential polynomial $f$ in variables $u_0,v_0,u_1,v_1,\ldots$  is a finite
linear combination of the form
\[ f=\sum_{(\alpha,\beta)\in A}a_{\alpha,\beta}u^\alpha v^\beta\, ,
\qquad a_{\alpha,\beta}\in\bbbc\, ,\] where the sum is taken over a finite set
$$A=\{(\alpha,\beta)\, |\, \alpha,\beta\in\zpy,\ |\alpha|+|\beta|>0\}.$$
It is a differential ring. We again denote it $\cR$. Derivation $D_x$, (cf.
(\ref{Dx})) is now replaced by
\[ D_x =\sum_{k\ge 0}\left(u_{k+1}\frac{\partial}{\partial u_k}+
v_{k+1}\frac{\partial}{\partial v_k}\right)\, . \] For any $f\in\cR$ the Fr\'echet
derivative $f_*$ is a (row) vector operator
\begin{equation}\label{sysfrechet}
f_*=(f_{*u},f_{*v})=\left(\sum_{k\ge 0}\frac{\partial f}{\partial u_k}D_x^k\ ,
\ \ \ \sum_{k\ge 0}\frac{\partial f}{\partial v_k}D_x^k\right)\, .
\end{equation}

Systems of two evolutionary equations we will write in vector form
\begin{equation}\label{evsysvec}
 \bu_t=\bF (\bu, \bu_1,\ldots ,\bu_n),
\end{equation}
where $\bu_k=(u_k,v_k)^T$ and $\bF=(F_1,F_2)^T$ are vector-columns where
$F_1,F_2\in\cR$ (the upper index $T$ stands for the transposition).

Let us introduce an infinite dimensional linear space over $\bbbc$
$$\cL=\{(H_1,H_2)^T\, |\, H_1,H_2\in\ring\}.$$  We equip $\cL$ with a  Lie bracket
\[ [\bF,\bG]=\bF_*(\bG)-\bG_*(\bF)\in\cL ,\]
where the Fr\'echet derivative $\bH_*$ for any $\bH\in\cL$ is defined as
\[ \bH_*=\left(\begin{array}{cc}
H_{1\, *u}&H_{1\, *v}\\
H_{2\, *u}&H_{2\, *v}
\end{array}\right)\, .
\]
Thus $\cL$ has a structure of an infinite dimensional Lie algebra over $\bbbc$.
Subalgebra of symmetries of equation (\ref{evsysvec}) is the centraliser $C_\cL(\bF)$
of $\bF$ in $\cL$ (cf. section \ref{diffpoly1}).

Evolutionary system (\ref{evsysvec}) defines
a derivation $D_\bF :\cR\mapsto\cR$
\[ D_\bF(a)=a_{*u}(F_1)+a_{*v}(F_2), \]
which is also a derivation of the Lie algebra $\cL$. This derivation we often denote
as $D_t$.  In this notations derivation  $D_x$ coincides with $D_{\bf u_1}$ and
$D_u+D_v$ with $D_{\bf u}$.

The ring $\cR$ has several gradings. Here we define a {\em monomial degree} grading
\[ \cR=\bigoplus_{k\in\bbbn}\cR^k\, ,\quad \cR^{k}=\{ a\in\cR\, |\, D_{\bf u}a=ka\}.
\]
We say that $a=o(\cR^n)$ if $a\in \bigoplus_{k>n}\cR^k$.

Lie algebra $\cL$ inherits the gradings of $\cR$. A {\em monomial degree} grading
\[ \cL=\bigoplus_{k\in\zp}\cL^k\, ,\quad \cL^{k}=\{ \bH\in\cL\, |\,
D_{\bf u}\bH=(k+1)\bH\},
\]
and thus $[\cL^p,\cL^q]\subset \cL^{p+q}$ is convenient for the definition of
approximate symmetries. For $\bH\in\cL$ we say that $\bH=o(\cL^n)$ if $\bH\in
\bigoplus_{k>n}\cL^k$. Approximate symmetries of equation (\ref{evsysvec}) of degree
$n$ are defined as elements of the approximate centraliser
\[C_\cL^n(\bF)=\{\bG\in\cL\, |\, [\bF,\bG]=o(\cL^{n-1})\},\]
which is a subalgebra of $\cL$.

The {\em weighted gradation} is useful for the study of homogeneous systems. We
assign weights ${\bf w}=(w_u, w_v)$ with rational entries to the vector variable
${\bf u}$ and define a linear operator
\[
W=(w_u D_u+w_v D_v+X_u+X_v)\left(\begin{array}{cc}
1&0\\0&1
\end{array}\right)-\left(\begin{array}{cc}
w_u&0\\0&w_v
\end{array}\right)\,
\]
with spectrum $S_{ W}=\{(p-1)w_u+(q-1)w_v+r\, |\, p,q,r\in\zp,\, p+q>0 \}$.
Then the linear subspaces $\cL_\mu$ in the decomposition
\begin{equation*}\label{w-grad}
 \cL=\bigoplus_{\mu\in S_W}\cL_\mu\, ,\quad \cL_\mu=\{\bH\in\cL\, |\,
W\bH=\mu\bH  \}
\end{equation*}
satisfy to
\begin{equation}\label{w-gradLie}
  [\cL_\mu,\cL_\nu]\subset\cL_{\mu+\nu}\, .
\end{equation}

Elements of $\cL_\mu$ we call ${\bf w}$--homogeneous differential polynomial vectors
of weight $\mu$.

For example, if the weight vector of variables $(u,v)$ is ${\bf w}=(1/2,1)$, then
\[
\bF=\left(\begin{array}{l}
 v_1\\
u_{2}+3uv_1+vu_1-3u^2u_1
\end{array}\right)
\]
is ${\bf w}$--homogeneous element of weight $3/2$, indeed $\bF\in\cL_{3/2}$.

If $\bF\in\cL_\mu$ is a homogeneous vector, then $D_\bF$ is a homogeneous derivation
of weight $\mu$:
\[D_\bF \cL_\nu\subset \cL_{\nu+\mu}\, ,\qquad \bF\in\cL_\mu.\]
From (\ref{w-gradLie}) it immediately follows
\begin{Pro}
Let $\bG=\bG_{\nu_1}+\cdots + \bG_{\nu_m}\, ,\ \bG_\gamma\in\cL_\gamma$ be a
generator of a symmetry of a homogeneous equation, then  each ${\bf w}$--homogeneous
component $\bG_{\nu_k}$  is a generator of a symmetry.
\end{Pro}

For evolutionary system (\ref{evsysvec}) a recursion operator (a formal recursion
operator)  $\Lambda$ can be defined as a differential or pseudo-differential operator
(or a formal series)
\[
\Lambda=\Lambda_k D_x^k+\Lambda_{k-1} D_x^{k-1}+\cdots,\quad \Lambda_s\in
\mbox{Mat}_{2\times 2}(\cR)\ ,
\]
 which satisfies
the following operator equation
\begin{equation}\label{recopsys}
D_{\bF} (\Lambda)={\bF}_* \circ \Lambda-\Lambda\circ {\bF}_* \,
\end{equation}
(compare with Definition \ref{defrecop1}). If action of $\Lambda$ is well defined on
a symmetry ${\bG}_1$, i.e. ${\bG_2}=\Lambda({\bG}_1)\in \cL$, then ${\bG}_2$ is a new
symmetry of the evolutionary system (\ref{evsysvec}). Starting from a ``seed''
symmetry ${\bG}_1$, one can build up an infinite hierarchy of symmetries
${\bG}_{n+1}=\Lambda^n({\bG}_1)$, provided that each action of $\Lambda$ produces an
element of  $\cL$.

Symbolic representation of the ring $\cR$ generated by two independent variables
$u,v$ is quite straightforward. It is a $\bbbc$--linear isomorphism which is
sufficient to define for the monomials. Suppose we have a monomial $u^\alpha
v^\beta$. Let the symbolic representation for the monomial  $u^\alpha$ be
$\hat{u}^{|\alpha|}a(\xi_1,\ldots ,\xi_{|\alpha|})$ where $a(\xi_1,\ldots
,\xi_{|\alpha|})$ is a symmetrical polynomial (see Definition \ref{repmon}). Acting
by the same rule, but reserving a set of variables $\zeta_1, \zeta_2, \ldots $
(instead of $\xi_1,\xi_2, \ldots $) for the symbolic representation of $v$--monomials
we get
 $v^\beta \longmapsto \hat{v}^{|\beta|} b(\zeta_1,\ldots ,\zeta_{|\beta|})$. Then
\[ u^\alpha v^\beta \longmapsto \hat{u}^{|\alpha|}\hat{v}^{|\beta|}a(\xi_1,\ldots,
\xi_{|\alpha|})b(\zeta_1,\ldots ,\zeta_{|\beta|})\, . \] Note that the symbol
obtained is invariant with respect to the direct product of two permutation groups
$\cS_{|\alpha|}^\xi \times \cS_{|\beta|}^\zeta$.

To the product of two elements $f,g\in\ring$ with symbols $$f\mapsto
\hat{u}^n\hat{v}^ma(\xi_1,\ldots,\xi_n,\zeta_1,\ldots,\zeta_m)\ \mbox{and}\ g\mapsto
\hat{u}^p\hat{v}^qb(\xi_1,\ldots,\xi_p,\zeta_1,\ldots,\zeta_q)$$ corresponds:
\begin{eqnarray}\label{multmonoms}
f g&\longmapsto& \hat{u}^{n+p}\hat{v}^{m+q}\langle\langle
a(\xi_1,\ldots,\xi_n,\zeta_1,\ldots,\zeta_m)\nonumber\\
&&b(\xi_{n+1},\ldots,\xi_{n+p},
\zeta_{m+1},\ldots,\zeta_{m+q})\rangle_{\cS^\xi_{n+p}}\rangle_{\cS^\zeta_{m+q}},
\end{eqnarray}
where the symmetrisation operation is taken with respect to permutations of all
arguments $\xi$ and then $\zeta$ (the symmetrisation can be made in any order).

If $f\in\ring$ has a symbol $f\longmapsto
\hat{u}^n\hat{v}^ma(\xi_1,\ldots,\xi_n,\zeta_1,\ldots,\zeta_m)$, then the
symbolic representation for the derivative $D_x(f)$ is:
\[
D_x(f)\longmapsto
\hat{u}^n\hat{v}^m(\xi_1+\cdots+\xi_n+\zeta_1+\cdots+\zeta_m)
a(\xi_1,\ldots,\xi_n,\zeta_1,\ldots,\zeta_m) .
\]

To the operator $D_x$ we shall assign a special symbol $\eta$
satisfying the following composition rule (the Leibnitz rule)
\begin{eqnarray*}
&&\eta\circ \hat{u}^n\hat{v}^m
a(\xi_1,\ldots,\xi_n,\zeta_1,\ldots,\zeta_m)\\
&=&\hat{u}^n\hat{v}^m a(\xi_1,\ldots,\xi_n,\zeta_1,\ldots,\zeta_m)(\xi_1+\cdots
+\xi_n+\zeta_1+\cdots +\zeta_m+\eta )\, .
\end{eqnarray*}
For $f\in\ring$ with symbol $f\longmapsto
\hat{u}^n\hat{v}^ma(\xi_1,\ldots,\xi_n,\zeta_1,\ldots,\zeta_m)$ the symbolic
representation of the Fr\'echet derivative (\ref{sysfrechet}) is
\begin{eqnarray*}
&&f_{*u}\longmapsto \hat{u}^{n-1}\hat{v}^m
na(\xi_1,\ldots,\xi_{n-1},\eta,\zeta_1,\ldots,\zeta_m),\\
&&f_{*v}\longmapsto
\hat{u}^n\hat{v}^{m-1}ma(\xi_1,\ldots,\xi_n,\zeta_1,\ldots,\zeta_{m-1},\eta)\, .
\end{eqnarray*}

\section{Integrability of evolutionary equations}\label{sec3}
\subsection{Study of symmetries of evolutionary equations in symbolic
representation}\label{sec24}

Using the above symbolic representation of the Lie bracket we can study the
properties of symmetries in great details.

\begin{The}\label{theorsym}
Let the right hand side of evolutionary equation (\ref{eveq1}) has symbolic
representation
\begin{equation*}\label{Fsymb}
F \longmapsto \hu\omega(\xi_1)+\hu^2a_1(\xi_1,\xi_2)
+\hu^3a_2(\xi_1,\xi_2,\xi_3)+\cdots
\end{equation*}
and the degree of polynomial $\omega(\xi_1)$ is greater than $1$. If
\begin{equation}\label{Gsymb}
 G\longmapsto \hu\Omega(\xi_1)+\hu^2A_1(\xi_1,\xi_2)
 +\hu^3A_2(\xi_1,\xi_2,\xi_3)+\cdots
\end{equation}
is a symmetry, then its coefficients can be found recursively
\begin{eqnarray}\label{sA1}
&&A_1(\xi_1,\xi_2)=\frac{G^\Omega (\xi_1,\xi_2)} {G^\omega
(\xi_1,\xi_2)}a_1(\xi_1,\xi_2)\\ \label{Am}
&&A_{m-1}(\xi_1,...,\xi_{m})=\frac{1}{G^\omega (\xi_1,...,\xi_{m})}\bigg(G^\Omega
(\xi_1,...,\xi_{m})a_{m-1}(\xi_1,...,\xi_{m})\nonumber\\
&&+ \sum_{j=1}^{m-2} \left\langle(j+1) A_j(\xi_1,...,\xi_j,\sum_{l=j}^{m-1}
\xi_{l+1})a_{m-1-j}(\xi_{j+1},...,\xi_{m})\right.
\\ \nonumber
&& \left. -(m-j) a_{m-1-j} (\xi_1,...,
\xi_{m-1-j},\sum_{l=0}^{j}\xi_{m-l})A_j(\xi_{m-j},\ldots,\xi_{m}) \right\rangle
_{\cS^{\xi}_{m}}\bigg),
\end{eqnarray}
where
\begin{equation}\label{Gomega}
G^\omega (\xi_1,...,\xi_m)=\omega(\sum_{n=1}^{m}\xi_n)-\sum_{n=1}^{m}\omega(\xi_n)\,
.
\end{equation}
\end{The}
{\bf Proof:} The proof of the Theorem is straightforward (see, for example
\cite{mr1908645}). Using (\ref{slie}) we can compute the Lie bracket between $F$ and
$G$. When it vanishes up to $\hring^2$, we express $A_1(\xi_1,\xi_2)$ from the
result, which leads to formula (\ref{sA1}). The Lie bracket vanishing up to
$\hring^{m}$ is equivalent to formula (\ref{Am}). \hfill$\square$

\begin{Cor}\label{cor1}
For the equation stated in Theorem \ref{theorsym}, (i). any symmetry has a linear
part, that is, $\Omega(\xi_1)\neq 0 $; (ii). algebra of symmetries is
commutative.
\end{Cor}
{\bf Proof:} (i). Let us assume that $\Omega (\xi_1)=0$. Then it follows from
(\ref{sA1}) that $A_1(\xi_1,\xi_2)=0$. Assuming that $A_k=0$ for all $1<k<m-1$, we
get from (\ref{Am}) that $A_{m-1}=0$. Thus by induction, we get $G=0$.

(ii). Commutator of two symmetries is a symmetry due to the Jacobi identity, but it
does not contain a linear part. Thus it must vanish. \hfill$\square$

Theorem \ref{theorsym} states that a symmetry of equation is uniquely determined by
its linear part (i.e. dispersion).  For fixed $\Omega(\xi_1)$ all coefficients in the
series (\ref{Gsymb}) can be found recursively. Theorem \ref{theorsym} does not mean
that any evolutionary equation has a symmetry. The right hand side of (\ref{Gsymb})
must represent a valid symbol, i.e., an element of $\hring$. Thus:
\begin{enumerate}
\item [a.] all coefficients $A_{m}(\xi_1, \ldots , \xi_{m+1})$ must be polynomial,
\item[b.] it should be a finite number of non-vanishing coefficients
$A_{m}$.
\end{enumerate}
In general, as it follows from (\ref{sA1}) and (\ref{Am}), the coefficients $A_k$ are
rational functions -- they have denominators $G^\omega $. In order to define a symbol
of a symmetry, these denominators must cancel with appropriate factors in the
numerators.  Thus factorisation properties of polynomials $G^\omega\) and \(G^\Omega$
are crucial for the structure of the symmetry algebra of the equation.

\begin{Pro}{\bf(F. Beukers \cite{Beuk97})}\label{Beuk}
For any positive integer \(m\ge 2\) the polynomial
\[h_{c,m}=(\xi_1+\xi_2+\xi_3+\xi_4)^m-c_1^{m-1} \xi_1^m-c_2^{m-1} \xi_2^m
-c_3^{m-1} \xi_3^m-c_4^{m-1} \xi_4^m,\]
where \(\Pi_{i=1}^4 c_i\neq 0\), is irreducible over \(\bbbc\).
\end{Pro}
{\bf Proof:} Suppose that \(h_{c,m}=A\cdot B\) with \(A,\) and \( B\) two polynomials
of positive degree. Then the projective hypersurface \(\Sigma\) given by
\(h_{c,m}=0\) consists of two components $\Sigma_A,\Sigma_B$ given by \(A=0, B=0\)
respectively. The intersection $\Sigma_A \bigcap\Sigma_B$ consists an infinite number
of points, which should be singularities of \(\Sigma\) since
\[ \left. \frac{d h_{c,m}}{d \xi_i}=\frac{d A}{d \xi_i} \cdot B\right|_{\Sigma_A
\bigcap\Sigma_B}\left.  +A \cdot \frac{d B}{d \xi_i}\right|_{\Sigma_A
\bigcap\Sigma_B} =0.\] Thus it suffices to show that \(\Sigma\) has finitely many
singular points.

We compute the singular points by setting the partial derivatives of \(h_{c,m}\)
equals to zero, i.e.,
\begin{eqnarray*}
\left\{\begin{array}{l}
(\xi_1+\xi_2+\xi_3+\xi_4)^{m-1}-(c_1 \xi_1)^{m-1}=0\\
(\xi_1+\xi_2+\xi_3+\xi_4)^{m-1}-(c_2 \xi_2)^{m-1}=0\\
(\xi_1+\xi_2+\xi_3+\xi_4)^{m-1}-(c_3 \xi_3)^{m-1}=0\\
(\xi_1+\xi_2+\xi_3+\xi_4)^{m-1}-(c_4 \xi_4)^{m-1}=0
\end{array}\right.
\end{eqnarray*}
From these equation follows in particular that
\[\xi_1= \zeta_1/c_1,\ \xi_2=\zeta_2/c_2,\ \xi_3=\zeta_3/c_3,\ \xi_4=\zeta_4/c_4,\]
where \(\zeta_i^{m-1}=1\) and \(\zeta_1/c_1 +\zeta_2/c_2 +
\zeta_3/c_3+\zeta_4/c_4=1\). For given \(c_i, i=1\cdots 4\), we get finitely many
singular points. \hfill$\square$

\begin{Cor}\label{lemma1}
Polynomials $G^\omega (\xi_1,\ldots,\xi_n)$, cf. (\ref{Gomega}) are irreducible for
$n\ge 4$.
\end{Cor}
{\bf Proof:} Let $\omega(\xi)=\alpha^m\xi^m+\cdots +\alpha_0$. If $0\le m\le1$ then
$G^\omega$ is a constant and therefore irreducible. If $m\ge 2$ polynomial $G^\omega$
has the form \[G^\omega(\xi_1,\ldots,\xi_n)=\alpha_m
G^{(m)}(\xi_1,\ldots,\xi_n)+g^\omega,\] where $\deg(g^\omega)<m$ and
\begin{eqnarray}\label{gfun}
G^{(m)}(\xi_1,\ldots,\xi_n)=(\xi_1+\cdots +\xi_n)^m-\xi_1^m-\cdots -\xi_n^m\, ,
\end{eqnarray}
which is irreducible according to Proposition \ref{Beuk}. \hfill $\square$

\begin{The}\label{TH2}
The algebra of symmetries of the evolutionary equation
\begin{equation}\label{eqprop}
u_t=\sum_{k=0}^n \alpha_k u_k+f(u_{n-1},\ldots , u)=F\, ,\quad n\ge 2,\ \ \alpha_n\ne 0
\end{equation}
where $f(u_{n-1},\ldots , u)\ne 0$ and
\begin{equation}\label{eqpropcond}
f(u_{n-1},\ldots , u)\in \bigoplus_{m>3}\, \,\bigoplus_{p<n}\cR^m_p
\end{equation}
is trivial, i.e., $C_\cR(F)=\spc\{ u_1,F\}$.
\end{The}
{\bf Proof:} In symbolic representation
\[ F \longmapsto  \hu \omega (\xi_1)+\hu^{m+1}a_{m}(\xi_1,\ldots ,\xi_{m+1})
+ \hu^{m+2}a_{m+1}(\xi_1,\ldots ,\xi_{m+2})+\cdots , \] where $\omega(\xi_1)=\alpha_n
\xi_1^n+\cdots+\alpha_1\xi_1 +\alpha_0$. The condition (\ref{eqpropcond}) implies
$m\ge 3$ and $\deg (a_{m}(\xi_1,\ldots ,\xi_{m+1}))<n$.

A symmetry of (\ref{eqprop}) is of the form
\[
 G\longmapsto \hu\Omega(\xi_1)+\hu^2A_1(\xi_1,\xi_2)
 +\hu^3A_2(\xi_1,\xi_2,\xi_3)+\hu^4A_3(\xi_1,\xi_2,\xi_3,\xi_4)+\cdots
\]
if it exists. We know that its linear part $\Omega(\xi_1)\neq 0$ from Corollary
\ref{cor1}. It follows from (\ref{Am}) that $A_k(\xi_1,\ldots ,\xi_{k+1})=0$ for
$k<m$ and
\begin{equation}\label{Amm} A_m(\xi_1,\ldots,\xi_{m+1})=\frac{G^\Omega
(\xi_1,...,\xi_{m+1})}{G^\omega(\xi_1,...,\xi_{m+1})}a_m(\xi_1,\ldots,\xi_{m+1}).
\end{equation}

Suppose $\Omega(\xi)\ne\alpha \xi+\beta \omega(\xi)$ for any $\alpha,\beta\in\bbbc$.
From Corollary \ref{lemma1}, we have that polynomials $G^\Omega (\xi_1,...,\xi_{m+1})$ and
$G^\omega(\xi_1,...,\xi_{m+1})$ are irreducible and therefore they are co-prime.
Since $$\deg (G^\omega(\xi_1,...,\xi_{m+1}))=n>\deg (a_{m}(\xi_1,\ldots
,\xi_{m+1})),$$ the right hand side of (\ref{Amm}) is a rational function (not a
polynomial). Thus there are no symmetries under the assumption.

When $\Omega(\xi)=\alpha \xi+\beta \omega(\xi)$ for some $ \alpha,\beta\in\bbbc$, it
follows from (\ref{Am}) that $G=\alpha u_1+\beta F\in \spc\{ u_1,F\}$.
\hfill$\square$

According to Definition \ref{Def3}, equation (\ref{eqprop}) is not integrable. In
(\ref{eqpropcond}) condition $p<n$  is essential. Indeed, equation
\[ u_t=u_2+u^m u_1^2 \]
($p=n=2$) is integrable for any $m$.

If an evolutionary equation (\ref{eveq1}) with linear
part of order $2$ or higher has a nontrivial symmetry, then any approximate symmetry
of degree $3$ is amendable to any degree. Thus, if we have infinitely many
approximate symmetries of degree $3$, then we have infinitely many approximate
symmetries of arbitrary high degree.

\begin{The}\label{MainR}
Let $\omega(\xi_1)$ be a polynomials of degree greater than $1$. Assume that
evolutionary equation (\ref{eveq1}) with linear terms $\hat{u}\omega(\xi_1)$ has a
nontrivial symmetry. Then for an approximate symmetry \( \sum_{j=1}^3 h_j,\ h_j \in
\cR^j\) of degree $3$, there exists a unique \( H=\sum_{j\ge 1} h_j \), \(h_j \in
\cR^j\) such that $H$ is an approximate symmetry of any degree.
\end{The}

This Theorem is the direct consequence of a more general Theorem 2.3 in
\cite{mr99g:35058} (see also Theorem 2.76 p.27 \cite{wang98}) formulated in the
context of filtered Lie modules. According to Theorem 2.3, in application to an
evolutionary equation (\ref{eveq1}) with linear terms $\hat{u}\omega(\xi_1)$, we
should require that polynomials $G^\omega (\xi_1,...,\xi_{m+1})$ and polynomials
$G^\Omega (\xi_1,...,\xi_{m+1})$ defined by (\ref{Gomega}) have no common factors for
some $m>1$. This is the case for $m=3$ as it follows from Corollary \ref{lemma1}.

The result of Theorem \ref{MainR} confirms the remark made in \cite{mr81i:35144}:
\begin{quote}{\em
Another interesting fact regarding the symmetry structure of evolution equations is
that in all known cases the existence of one generalised symmetry implies the
existence of infinitely many. (However, this has not been proved in general.)}
\end{quote}
For systems of equations and for non-evolutionary equations the conjecture that {\em
the existence of one symmetry implies the existence of (infinitely many) others} has
been disproved. In \cite{mr99i:35005} it has shown that example given in \cite{Ba91}
is indeed a counterexample to the conjecture (see also \cite{mnw05}). Even a
rectified conjecture \cite{fok87} that {\em for $N$-component equations one needs $N$
symmetries} is incorrect either. An example of a system of two equations possessing
exactly two nontrivial symmetries is given in  \cite{mr2003j:37109}.

These examples do not contradict to the spirit of our Theorem \ref{MainR} since they are
based upon the nonexistence of approximate symmetries of degree $2$, which is one of
the conditions in the theorem.

As we have already mentioned above, the existence of a formal recursion operator
$\Lambda$ (\ref{Lm}) for an evolutionary equation is a necessary condition for the
existence of an infinite hierarchy of symmetries. A similar, but not equivalent,
theorem can be stated in the symbolic representation. The difference is in the natural
ordering. In the standard representation the coefficients $l_k$  are ordered due to
the power of $D_x^k$ in the formal series $\Lambda$ (\ref{Lm}). In the symbolic
representation the natural ordering is due to the power of symbol $\hu$. The fact
that $l_k$ must be local, i.e. $l_k\in\ring$ in the symbolic representation suggests
the following definition:

\begin{Def}\label{local}
We say that function $b_m (\xi_1,...,\xi_m,\eta), \ m\ge 1$ is $k$-local if in the
expansion as $\eta\to\infty$
\[
b_m (\xi_1,...,\xi_m,\eta)=\beta_{m1}(\xi_1,...,\xi_m)\eta^{n_m}
+\beta_{m2}(\xi_1,...,\xi_m)\eta^{n_m-1}+ \, \cdots
\]
the first $k$ coefficients $\beta_{ms} (\xi_1,...,\xi_m),\ s=1,..., k$ are symmetric
polynomials in $\xi_1,...,\xi_m$. We say that  $b_m (\xi_1,...,\xi_m,\eta)$ is local
if it is $k$-local for any $k$.
\end{Def}

Existence of an infinite hierarchy of symmetries implies the existence of a formal
recursion operator with local coefficients (Proposition 3 in \cite{mr1908645}).
Existence of an infinite hierarchy of approximate symmetries of degree $N$ implies
that first $N-1$ coefficients of the formal recursion operator are local. The details
of the proof of the following Theorem one can find in \cite{mr1908645} (Proposition
3).

\begin{The}\label{theor4}
Suppose equation (\ref{eveq1}) has an infinite hierarchy of
approximate symmetries of degree $N$
\begin{equation*}
\label{symi} u_{t_i}=\hu\Omega_i(\xi_1)+\sum_{j\ge
1}\hu^{j+1}A_{ij}(\xi_1,\ldots,\xi_{j+1})=G_i\, ,\quad i=1,2,\ldots
\end{equation*}
where $\Omega_i(\xi _1)$ are polynomials of degree $m_i$ and $m_1 <m_2<\cdots <
m_{i}<\cdots $. Then the coefficients $\phi_m(\xi_1,...,\xi_m,\eta),\ m=1,... , N-1$
of the formal recursion operator
\begin{equation*}
\label{lp} \Lambda=\eta+\hu\phi_1(\xi_1,\eta )+\hu^2\phi_2(\xi_1,\xi_2,\eta )+\cdots
\end{equation*}
are local.
\end{The}

In symbolic representation equation (\ref{Leq}) can be solved \cite{mr1908645} in the
sense that coefficients of a formal series $\Lambda$ can be recursively for any
evolutionary equation (\ref{eveq1}):

\begin{The} \label{prolambda} Let $\phi(\eta )$ be an arbitrary function and
formal series
\begin{equation*}
\label{Lsym} \Lambda=\phi(\eta )+\hu\phi_1 (\xi_1,\eta )+\hu^2\phi_2
(\xi_1,\xi_2,\eta )+\hu^3\phi_3 (\xi_1,\xi_2,\xi_3,\eta )+\cdots
\end{equation*}
be a solution of equation (\ref{Leq}), then its coefficients
$\phi_m(\xi_1,...,\xi_m,\eta )$ can be found recursively
\begin{eqnarray*}
&&\phi_1 (\xi_1,\eta )=\frac{2(\phi(\eta+\xi_1)-\phi(\eta ))}{G^\omega(\xi_1,\eta )}
a_1(\xi_1,\eta )
\\   &&\phi_m
(\xi_1,...,\xi_m,\eta )=\frac{1}{G^\omega (\xi_1,...,\xi_m,\eta )} \bigg((m+1)
(\phi(\eta+\xi_1+... +\xi_m)\\
&& -\phi(\eta ))a_m(\xi_1,...,\xi_m,\eta )\\&&+\sum_{n=1}^{m-1}\langle n
\phi_n(\xi_1,..,\xi_{n-1},\xi_n+\cdots +\xi_m,\eta ) a_{m-n}(\xi_n,..,\xi_m)\\
 && +(m-n+1) \phi_n(\xi_1,..,\xi_{n},\eta+\sum_{l=n+1}^m\xi_{l})
a_{m-n}(\xi_{n+1},..,\xi_m,\eta )\\  &&
-(m-n+1)a_{m-n}(\xi_{n+1},..,\xi_m,\eta+\sum_{l=1}^n\xi_{l})
\phi_n(\xi_{1},..,\xi_n,\eta ) \rangle_{\cS^{\xi}_{m}} \bigg).
\end{eqnarray*}
\end{The}

Existence of formal recursion operator with local coefficients is a necessary
condition for the existence of an infinite hierarchy of symmetries. It suggests the
following test for integrability of equations  (\ref{eveq1}):
\begin{itemize}
\item Find a first few coefficients $\phi_n(\xi_1,...,\xi_n,\eta )$
(first three nontrivial coefficients $\phi_n$ were sufficient to analyse in all known
cases to us).
\item Expand these coefficients in series of $1/\eta$
\begin{equation*}\label{Phi}
\phi_n(\xi_1,...,\xi_n,\eta
)=\sum_{s=s_n}\Phi_{ns}(\xi_1,...,\xi_n)\eta^{-s}
\end{equation*}
and check that functions $\Phi_{ns}(\xi_1,...,\xi_n)$ are polynomials (not rational
functions).
\end{itemize}
This test will be extended and used for non-local and non-evolutionary equations in
sections \ref{bo-sec} and \ref{ch-sec}.

\subsection{Global classification of integrable homogeneous evolutionary equations}\label{sec_glob}
In this section, we give ultimate global classification of integrable equations of
the form
\begin{eqnarray}\label{evo}
u_t=u_n + f(u,\cdots,u_{n-1}),\qquad n\ge 2
\end{eqnarray}
where \(u_n + f(u,\cdots,u_{n-1})\) is a $\lambda$--homogeneous differential
polynomial and $\lambda\ge 0$. We give a complete description of integrable
equations for all $n$.

Theorem \ref{MainR} implies that if equation (\ref{evo}) possesses one higher
symmetry and infinitely many approximate symmetries of degree $3$, then it possesses
infinitely many approximate symmetries of any degree. Therefore to classify
integrable equations (\ref{evo}) it suffices to classify equations, which possess
infinitely many degree $3$ approximate symmetries and then impose the condition of
existence of at least one exact symmetry.  The classification has been done in the
case of $\lambda$-homogeneous equations with $\lambda\ge  0$. In the case $\lambda>0$
see the details in \cite{mr99g:35058}, while in the case $\lambda=0$ the details can
be found in \cite{mr2001h:37147}.

Now we sketch the results for the case $\lambda>0$ without the detailed proofs. The
following statement  on factorisation properties of polynomials
$G^{(k)}(\xi_1,\ldots,\xi_n)$ (\ref{gfun}) plays an important role in the classification
of integrable equations:
\begin{The}\label{BeukId}
\(G^{(k)}(\xi_1,\ldots,\xi_n)=t^{(k)} g^{(k)}\), where \((g^{(k)},g^{(l)})=1\) for
all \( k<l\), and \(t^{(k)}\) is one of the following cases.
\begin{itemize}
\item \(n = 2\):
\begin{itemize}
\item \(k=0\pmod 2\): \(\xi_1\xi_2\)
\item \(k=3\pmod 6\): \(\xi_1\xi_2(\xi_1+\xi_2)\)
\item \(k=5\pmod 6\): \(\xi_1\xi_2(\xi_1+\xi_2)(\xi_1^2+\xi_1\xi_2+\xi_2^2) \)
\item \(k=1\pmod 6\): \(\xi_1\xi_2(\xi_1+\xi_2)(\xi_1^2+\xi_1\xi_2+\xi_2^2)^2\)
\end{itemize}
\item \(n = 3\):
\begin{itemize}
\item \(k=0\pmod 2\): \(1\)
\item \(k=1\pmod 2\):
\((\xi_1+\xi_2)(\xi_1 +\xi_3) (\xi_2 +\xi_3)\)
\end{itemize}
\item \(n > 3\): \(1\)
\end{itemize}
\end{The}
For $n>3$ the statement follows from more general Theorem \ref{Beuk}. For $n=3$, it
has been proven by Beukers and was published in \cite{wang98,mr99g:35058} with his
kind permission.  The case $n=2$ has a quite remarkable history. In affine
co--ordinate $x=\xi_1/\xi_2$ we have $G^{(k)}(\xi_1,\xi_2)=\xi_2^k P_k(x)$ and the
problem is reducing to factorisation properties of the Cauchy-Liouville-Mirimanoff
polynomials
\[ P_k(x)=(1+x)^k-x^k-1\, . \]
The common factors $P_k=x(1+x)^\alpha (1+x+x^2)^\beta g^{(k)}(x)$ and their
periodicity have been established in the joint report of Cauchy and Liouville
\cite{cauchy-liouville}. Using Diophantine approximation theory Beukers has shown
that factors $g^{(k)},g^{(m)}$ are co-prime for $k\ne m$ \cite{mr98e:11029}. Beukers
also conjectured that factors $g^{(p)}$ are irreducible over $\bbbq$. For prime $p$
the irreducibility of $g^{(p)}$ over $\bbbq$ was earlier conjectured by Mirimanoff
\cite{Mirimanoff}. A remarkable progress towards the proof of the Mirimanoff
conjecture has been recently achieved in \cite{MR2317451}.

We now consider $\lambda$-homogeneous equations of the form
\begin{eqnarray}
&&u_t=u_n+f_2+f_3+\cdots,\quad f_i\in \cR^i\label{hoeq}\\
&\longmapsto & \hu a_0(\xi_1)+\hu^2a_1(\xi_1,\xi_2)
+\hu^3a_2(\xi_1,\xi_2,\xi_3)+\cdots,\nonumber
\end{eqnarray}
where $n\ge 2$, $\lambda>0$ and the degree of a polynomial $a_j$ is $n-j\lambda$. Note
that if \(\lambda\) is not integer and $i\lambda \notin {\bbbn}$, then \(a_{i}=0\).
This reduces the number of relevant \(\lambda\) to a
finite set.

Let \(G\in {\cR}\) be a nontrivial symmetry  of (\ref{hoeq}). Then it is of the form
\begin{eqnarray*}
&&G=u_m+g_2+g_3+\cdots,\quad g_i\in \cR^i\\&\longmapsto & \hu
A_0(\xi_1)+\hu^2A_1(\xi_1,\xi_2) +\hu^3A_2(\xi_1,\xi_2,\xi_3)+\cdots,
\end{eqnarray*}
where $2\le m\neq n$ and the degree of polynomial $A_j$ is $m-j\lambda$. For all
integers \( r \ge 0\) the following formula holds
\begin{eqnarray}\label{Condition}
\sum_{i=0}^r[ \hu^{i+1} a_{i},  \hu^{r-i+1} A_{r-i}]=0.
\end{eqnarray}
Clearly we have \( [ \hu a_{0},  \hu A_{0}]=0\). The next equation to be solved is
\([\hu a_0, \hu^2 A_1]+ [\hu^2 a_{1}, \hu A_{0}]=0\), which is trivially satisfied if
equation (\ref{hoeq}) has no quadratic terms: \(f_{2} = 0\). Let us concentrate on
the case \(f_{2}\neq 0\). In this case, using Theorem \ref{MainR}, we see that
the existence of a symmetry is uniquely determined by the existence of its quadratic
term \cite{wang98,mr99g:35058}.

We now make a very interesting observation. Assume $n$ and $q$ are both odd. Let us
compute the symmetry of equation (\ref{hoeq}) with linear term $u_{q}$. Its quadratic
terms, cf. (\ref{Condition}), have the following symbolic expression
\begin{eqnarray}\label{qua}
\frac{ a_{1}\ (\xi_1^2+\xi_1 \xi_2 + \xi_2^2)^{s-s'}\ g^{(q)}(\xi_1,\xi_2)}
{g^{(n)}(\xi_1,\xi_2)}.
\end{eqnarray}
Proposition \ref{BeukId} implies that \( \lambda \leq 3 +2\min(s,s')\), where
\(s'=\frac{n+3}{2}\pmod 3\) and \(s=\frac{q+3}{2}\pmod 3\). We see that if expression
(\ref{qua}) is a polynomial, then it defines a symmetry \( Q=u_q+Q_2+\cdots \) since
$Q$ is determined by its quadratic term $Q_2$. The evolutionary equations defined by
\(Q\) has the same symmetries as equation (\ref{hoeq}). So instead of (\ref{hoeq}) we
may consider the equation given by \(Q\). The lowest possible $q$ is  \( 2 s +3 \)
for \(s=0,1,2\). Therefore we only need to consider \(\lambda\)-homogeneous equations
with \(\lambda \leq 7\)) of orders \(\leq 7\).

A similar observation can be made for even \(n>2\). Suppose we have found a
nontrivial symmetry with quadratic term
\[
\frac{a_1\ G^{(q)}}{\xi_1\xi_2 \,g^{(n)}}
\]
This immediately implies \(\lambda \leq 2\). Then the quadratic term \( 2
\frac{a_1}{ g^{(n)}} \) defines a symmetry \( Q \) starting with $u_2$.
Therefore, we only need to find the symmetries of equations of order \(2\) to get the
complete classification of symmetries of \(\lambda\)-homogeneous scalar polynomial
equations (with \(\lambda \leq 2\)) starting with an even linear term.

Finally, we analyse the case when equation (\ref{hoeq}) has no quadratic terms. If
\(a_i=0\) for \(i=1,\cdots,j-1\), then we look at the equation \([ \hu a_{0},
\hu^{j+1} A_j]+ [\hu^{j+1} a_{j}, \hu A_{0}]=0\), i.e.
\begin{eqnarray*}\label{Srel}
A_{j}=\frac{G^{(m)}(\xi_1,\cdots, \xi_{j+1})\ a_{j}}
 {G^{(n)}(\xi_1,\cdots, \xi_{j+1}) } .
\end{eqnarray*}
From Proposition \ref{BeukId} and the proof of Theorem \ref{TH2}, we know there are
no symmetries for the equation when \(j\geq 3\), or when \(j=2\) and \(n\) is even.
When \(j=2\) and \(n\) is odd, it can only have odd order symmetries. In this case
one can remark that if the equation possesses symmetries for any $m$ then it must
possess a symmetry of order $3$.

By now, we have proved the following
\begin{The}\label{ReductionTheorem}
A nontrivial symmetry of a \( \lambda \)-homogeneous equation with $\lambda>0$ is
part of a hierarchy starting at order \(2,\ 3,\ 5\) or \(7\).
\end{The}
Only an equation with nonzero quadratic or cubic terms can have a nontrivial
symmetry. For each possible \(\lambda>0\), we must find a third order symmetry for a
second order equation, a fifth order symmetry for a third order equation, a seventh
order symmetry for a fifth order equation with quadratic terms, and the thirteenth
order symmetry for a seventh order equation with quadratic terms. The last case can
be easily reduced to the case of fifth order equations by determining the quadratic
terms of the equation. The details of this final computation are described
in \cite{mr99j:65233}.

\begin{The}\label{globclass} Let $\lambda> 0$. Suppose that a $\lambda$--homogeneous polynomial evolutionary
equation
$$ u_t=u_n + f(u,\cdots,u_{n-1}), \qquad n\ge 2$$
possesses nontrivial symmetries. Then it is a symmetry of one of the following
equations up to  a transformations \(u\mapsto \alpha u\), $\alpha \in \mathbb{C}$:

\noindent {\bf Burgers equation}
\[
u_t=u_2 +u u_1
\]
{\bf Korteweg--de Vries equation}
\[
u_t=u_3 +u u_1
\]
{\bf Potential Korteweg--de Vries equation}
\[
u_t=u_3 +u_1^2
\]
{\bf Modified Korteweg--de Vries equation}
\[
u_t=u_3 +u^2 u_1
\]
{\bf Ibragimov-Shabat equation}
\[
u_t=u_3 +3u^2 u_2 +9 uu_1^2 +3 u^4 u_1
\]
{\bf Kaup-Kupershmidt equation}
\[
u_t=u_5 +10 uu_3 +25 u_1u_2 +20 u^2u_1
\]
{\bf Potential Kaup-Kupershmidt equation}
\[
u_t=u_5 +10 u_1u_3 +\frac{15}{2} u_2^2 +\frac {20}{3}u_1^3
\]
{\bf Sawada-Kotera equation}
\[
u_t=u_5 +10 uu_3 +10 u_1u_2 +20 u^2 u_1
\]
{\bf Potential Sawada-Kotera equation}
\[
u_t=u_5 +10 u_1u_3 +\frac {20}{3} u_1^3
\]
{\bf Kupershmidt Equation }
\[
u_t=u_5 +5u_1u_3 +5u_2^2 -5u^2u_3 -20\,u u_1 u_2 -5u_1^3 +5 u^4u_1
\]
\end{The}

Finally we note that all the consideration in this section can be extended to the
case when the dependent variable $u$ takes its values in some associative
non-commutative algebra (such as matrix, operator, Clifford, and group algebras). A
complete classification for \(\lambda>0\) homogeneous equations with linear leading
term in the case of non-commutative free associative algebra was carried out in
\cite{mr1781148}.

\section{Classification results for Non-local integrable equations}\label{sec_nonlocal}
The perturbative symmetry approach in the symbolic representation allows to derive
integrability conditions for certain type of non-local equations. In this section we
consider two types of such equations: the Benjamin-Ono type and the Camassa-Holm type
\cite{mr1908645,mn2}.

Benjamin-Ono equation reads
\begin{equation}\label{BO}
u_t=H(u_2)+2uu_1,
\end{equation}
where $H$ denotes the Hilbert transform
\begin{equation*}
\label{hilbert} H(f)=\frac{1}{\pi}\int_{-\infty}^{\infty}\frac{f(y)}{y-x}dy.
\end{equation*}
It is well known that the higher symmetries and conservation laws of the Benjamin-Ono
equation contain nested Hilbert transform and thus  an appropriate
extension of the differential ring $\ring$ is required. The construction of such
extension is similar to the one proposed by Mikhailov and Yamilov in \cite{mr1643816}
for $2+1$ dimensional equations (see also section \ref{2+1}).

The second example is  Camassa-Holm type equation:
\begin{equation}
\label{CH}
m_t=cmu_1+um_1,\quad m=u-u_2,\quad c\in\bbbc\setminus\{ 0\}.
\end{equation}
This equation is known to be integrable for $c=2$ \cite{CamHolm} and  for $c=3$
\cite{DegPro}. Equation (\ref{CH}) is not in the evolutionary form, but if we exclude
one of the dependent variables (say $u$) then we obtain a non-local equation
\begin{equation}
\label{CHe}
m_t=cm\Delta(m_1)+\Delta(m)m_1,\quad \Delta=(1-D_x^2)^{-1}
\end{equation}
and again the ring extension is required.

The symbolic representation and the concept of quasi-locality \cite{mr1643816} are
the key ideas in the extension of the symmetry approach to non-local equations. In
the definitions of all basic objects such as symmetries, formal recursion operators,
conservation laws  etc. we replace the ring of differential polynomials $\ring$ by an
appropriately extended ring. Elements of this extended ring we call quasi-local
polynomials (see details in the next section). Symbolic representation gives us a
simple criteria to decide if a given expression belong to the extended ring. In this
extended setting Theorem \ref{theorsym} and most of the results of Section
\ref{sec24} hold if we just replace ``local" by ``quasi-local" in conditions and
statements.

\subsection{Benjamin-Ono type equations and  Intermediate long wave equation}\label{bo-sec}

Let us consider the following sequence of ring extensions:
\begin{equation*}\label{R^n}
{\cal R}_{H^0}=\ring\, , \quad {\cal R}_{H^{n+1}}=\overline{{\cal R}_{H^n}\bigcup
H({\cal R}_{H^n})}\, ,
\end{equation*}
where the set $H({\cal R}_{H^n})$ is defined as $H({\cal R}_{H^n})=\{
H(a); a\in{\cal R}_{H^n}\}$ and the horizontal line denotes the ring
closure. Each ${\cal R}_{H^n}$ is a ring and the index $n$
indicates the nesting depth of the operator $H$:
\[
{\cal R}_{H^0}\subset{\cal R}_{H^1}\subset{\cal
R}_{H^2}\subset\cdots\subset{\cal R}_{H^n}\subset\cdots\subset {\cal R}_{H^\infty}={\cal
R}_H.
\]
Elements of ${\cal R}_{H^n},\,  n\ge 1$ we call {\sl quasi--local polynomials}. The
right hand side of equation (\ref{BO}), its symmetries and densities of conservation
laws are quasi--local polynomials.

We now consider scalar evolutionary equations, whose right hand side is a quasi--local
polynomial
\begin{equation}
\label{heq} u_t=F,\quad F\in{\cal R}_H.
\end{equation}
For the definition of its symmetry we replace $\ring$  by ${\cal R}_H$ in Definition
\ref{def_sym}. Actual computations in ${\cal R}_H $ lead to quite cumbersome
calculations. On the other hand, in the symbolic representation computations simplify
drastically and results can be neatly formulated.

In the symbolic representation operator $H$ is represented by $i\sign (\eta )$. So
the symbolic representation of the ring extensions is obvious.  Suppose $f\in {\cal
R}_{H^0}$ and
\[
f\mapsto \hat{u}^n a(\xi_1,\ldots,\xi_n).
\]
Then
\[
H(f)\mapsto \hat{u}^n i\sign(\xi_1+\cdots+\xi_n)a(\xi_1,\ldots,\xi_n).
\]
In the extended ring all the definitions, such as the Fr\'echet derivative, Lie
bracket and approximate symmetries, are exactly the same as in the local case.
However, the symbols of elements of the extended ring are symmetric sign-polynomials
instead of symmetric polynomials. For example, the symbolic representation of
$H(u_n)$ and $H(uH(u_1))$ is:
\begin{eqnarray*}
&&H(u_n)\mapsto \hat{u}\, i\sign(\xi_1)\xi_1^n,\\
&& H(uH(u_1))\mapsto -\frac{\hat{u}^2}{2}\sign(\xi_1+\xi_2)
\left(\xi_1\sign(\xi_1)+\xi_2\sign(\xi_2)\right).
\end{eqnarray*}
The symbolic representation of the Benjamin-Ono equation (\ref{BO})
is
\begin{equation*}
\label{BOs} u_t=i \hat{u} \sign
(\xi_1)\xi_1^2+\hat{u}^2(\xi_1+\xi_2).
\end{equation*}
Counting the degrees of sign-polynomials we assume that
$\deg(\sign(\xi_1+\cdots+\xi_k))=0$.

Theorem \ref{theorsym} for evolutionary equations and symmetries in $\ring_H$, in the
symbolic representation  remains the same with the only amendment that all symbols
now are sign--polynomials. To introduce  a formal recursion operator for equation
(\ref{heq}) we introduce a notion of asymptotically local functions, which
generalises the notion of local functions, cf. Definition \ref{local}:

\begin{Def} A function $a_n(\xi_1,\ldots,\xi_n,\eta)$ is called
asymptotically local if the coefficients $a_{np}(\xi_1,\ldots,\xi_n)$ and
$\tilde{a}_{np}(\xi_1,\ldots,\xi_n)$ of its expansion at $\eta\to\infty$:
\[
a_{n}(\xi_1,..,\xi_n,\eta)=\sum_{p=s_{n}}^{\infty}a_{np}(\xi_1,..,\xi_n)\eta^{-p}+
\sum_{p=\tilde{s}_{n}}^{\infty}\tilde{a}_{np}(\xi_1,..,\xi_n)\sign(\eta)\eta^{-p}
\]
are sign--polynomials, i.e. represent elements from the extended ring $\ring_H$.
\end{Def}

In the above  expansion we take into account $\sign(\eta+\sum_j\xi_j)=\sign(\eta)$ as
$\eta\to\infty$.

\begin{Def} A formal series
\[
\Lambda=\phi(\eta)+\hat{u}\phi_1(\xi_1,\eta)+\hat{u}^2\phi_2(\xi_1,\xi_2,\eta)
+\hat{u}^3\phi_3(\xi_1,\xi_2,\xi_3,\eta)+\cdots
\]
is called a formal recursion operator of equation (\ref{heq}) if it
satisfies equation (\ref{Leq}) and all its coefficients are
asymptotically local.
\end{Def}
Without loss of generality function $\phi(\eta)$ can be chosen as
either $\phi(\eta)=\eta$ or $\phi(\eta)=\eta\sign(\eta)$.

As in the local case, we can solve equation (\ref{Leq}) with respect
to coefficients of the formal recursion operator and Proposition \ref{prolambda}
holds.
The  generalisation of Theorem \ref{theor4} is straightforward,
we just replace ``local'' by asymptotically local.

For the Benjamin-Ono equation (\ref{BO}), the first coefficient $\phi_1(\xi_1,\eta )$
of corresponding formal recursion operator $$\Lambda=\eta+u\phi_1(\xi_1,\eta
)+u^2\phi_2(\xi_1,\xi_2,\eta )+\cdots$$ looks like
\[
\phi_1(\xi_1,\eta )=\sign(\eta )+\frac{\xi_1(\sign(\xi_1)+\sign(\eta
))}{2\eta}+O(\frac{1}{\eta^7})
\]
and it is asymptotically local. One may easily check asymptotic locality of other
coefficients $\phi_2(\xi_1,\xi_2,\eta )$, $\phi_3(\xi_1,\xi_2,\xi_3,\eta ),...$.

In this setting we can classify the generalisation of Benjamin-Ono
type equation. Consider an equation of the form
\begin{eqnarray}
u_t&=&H(u_2)+c_1uu_1+c_2H(uu_1)+c_3uH(u_1)+c_4u_1H(u)+\nonumber\\
&& +c_5H(u H(u_1))+ c_6H(u){H}(u_1),\label{BOg}
\end{eqnarray}
where $c_j$ are complex constants. The linear term of this equation
 coincides with the linear term of the
Benjamin-Ono equation and all possible homogeneous terms are
included if we suppose that $H$ is a zero-weighed operator
$W(H(f))=W(f)$ and weight of the variable $u$ equals $2$. We also
take into account that ${H}^2=-1$ and the Hilbert-Leibnitz rule
\[
{H}(fg)=f{H}(g)+g{H}(f)+{H}({H}(f){H}(g)).
\]

The following theorem holds (see the details and proof  in \cite{mn2}).
\begin{The}
Equation of the form (\ref{BOg}) possesses an infinite hierarchy of higher symmetries
if and only if it is, up to the point transformation $u\mapsto a u+b {H}(u)$,
$a^2+b^2\ne 0$ and re-scalings $x\mapsto\alpha x$, $t\mapsto\beta t$,
$a,b,\alpha,\beta\in\bbbc$, one of the list
\begin{eqnarray}
u_t&=&{H}(u_2)+D_x(\frac{1}{2}c_1u^2+c_2u{H}(u)-\frac{1}{2}c_1(u)^2);\label{eqA}\\
u_t&=&{H}(u_2)+D_x(\frac{1}{2}c_1u^2+\frac{1}{2}c_2{H}(u^2)-c_2u{H}(u));\label{eqB}\\
u_t&=&{H}(u_2)+i uu_1\pm {H}(uu_1)\mp u {H}(u_1)\mp2u_1{H}(u)\nonumber\\
&&-i {H}(u {H}(u_1));\label{eqC}\\
u_t&=&{H}(u_2)+{H}(uu_1)+u_1{H}(u)\pm i {H}(u {H}(u_1))\nonumber\\
&&\pm i {H}(u){H}(u_1).\label{eqD}
\end{eqnarray}
\end{The}
The proof of this theorem requires to check the quasi--locality of the first three
coefficients of the corresponding formal recursion operators. Equations  (\ref{eqA}),
(\ref{eqC}) and (\ref{eqD}) can be reduced to the Burgers equation. In the case
$c_1^2+c_2^2\ne 0$ equation (\ref{eqB}) can be transformed into the Benjamin-Ono
equation (\ref{BO}). When $c_1^2+c_2^2= 0$, it is equivalent to the Burgers equation.
The explicit form of transformations are in \cite{mn2}. The properties of the
Benjamin-Ono equation have been studied in \cite{AS}.

Finally we draw our attention to the intermediate long wave  equation
\begin{equation*}
\label{ILW} u_t=-\delta^{-1}u_1+2uu_1+{\cal T}(u_2),
\end{equation*}
where $\delta $ is a real constant parameter and 
\[ {\cal T}(u(x)):=\frac{1}{2\delta}\int_{-\infty}^{\infty}
\coth\left(\frac{\pi}{2\delta}(x-y)\right)u(y)dy\ .\] 
This equation was derived by
Joseph \cite{Jos} as the equation describing propagation of non-linear waves in a
fluid of finite depth. The intermediate long wave equation is an intermediate between Benjamin-Ono and
the Korteweg--de Vries  equations in the sense that the limit $\delta\to\infty$
yields Benjamin-Ono equation, while $\delta\to 0$ gives KdV equation. The intermediate long wave
equation possesses an infinite hierarchy of higher symmetries and is integrable by
the inverse scattering method \cite{AFSS}. As in the case of the Benjamin-Ono
equation, all its higher symmetries contain nested ${\cal T}$ operator.

We consider the general non-linear equation of the intermediate long wave form with some linear operator
${\cal T}$
\begin{equation}
\label{ILWg}
u_t={\cal T}(u_2)+2uu_1
\end{equation}
and address the question: for which linear operators ${\cal T}$ this equation
possesses an infinite hierarchy of higher symmetries/conservation laws? In
\cite{AFSS} Ablowitz {\it et al} have shown that if equation (\ref{ILWg}) possesses
infinitely many conservation laws then the conditions
\begin{equation}
\label{Tc}
{\cal T}(u{\cal T}v+v{\cal T}u)=({\cal T}u)({\cal T}v)-uv,
\end{equation}
\begin{equation*}\label{Tc1}
\int_{-\infty}^{\infty}(u{\cal T}v+v{\cal T}u)dx=0
\end{equation*}
must be satisfied.

The perturbative symmetry approach  allows to derive conditions for the operator
${\cal T}$ necessary for  the existence of an infinite hierarchy of higher
symmetries. All the steps are similar to the case of Benjamin-Ono type equation:
\begin{itemize}
\item Extend the differential ring by the operator ${\cal T}$ exactly in the same way
as we did with $H$ and define $\ring _{\cal T}$ and its symbolic representation.
\item Define  higher symmetries in the extended ring $\ring_{\cal T}$.
\item Introduce a formal recursion operator and asymptotic locality of its coefficients.
\end{itemize}

Without going into the details (see \cite{mr1908645} and \cite{HN}) we present the
following statement:
\begin{The} Assume that the operator ${\cal T}$ has the symbolic representation $i f(k)$:
\[
{\cal T}(u(x))=i\int_{-\infty}^{\infty}f(k)\hat{u}(k)e^{ikx}dk
\]
and that $f(k)\to 1$, faster than any power of $k^{-1}$, as $k\to +\infty$. Then if
equation (\ref{ILWg}) possesses a formal recursion operator with first three
coefficients being asymptotically local then $f(k)$ satisfies the functional
equation:
\begin{equation}\label{feq}
f(x+y)\left(f(x)+f(y)\right)=f(x)f(y)+1
\end{equation}
\end{The}

Formula (\ref{feq}) is equivalent to (\ref{Tc}) in the symbolic representation. Its
general odd solution, smooth on the real line except the origin, is given by
\[
f(k)=\coth(\delta k)
\]
which corresponds to the intermediate long wave equation and the limiting case $\delta\to +\infty$
corresponds the Benjamin-Ono equation
\[
f(k)=\sign(k).
\]
These are the only such equations possessing infinitely many conservation laws.  The
only even solution of (\ref{feq}) is
$$
f(k)=const
$$
leading to the Burgers equation (up to a re-scaling), which has no non-trivial
conservation laws.

\subsection{Camassa-Holm type equations}\label{ch-sec}
We now apply the perturbative symmetry approach to determine integrable cases of the
Camassa-Holm type equation (\ref{CHe}). It contains operator $\Delta=(1-D_x^2)^{-1}$
and therefore we extend the differential ring $\ring$ by operator $\Delta$ and define
the $\Delta$--extended ring $\ring_{\Delta}$ as we did for $\ring_H$ in section
\ref{bo-sec}.

The symbolic representation of the operator $\Delta$ is
$\Delta\mapsto\frac{1}{1-\eta^2}$. Therefore if $f\in \ring$ with symbol
$\hat{u}^na(\xi_1,\ldots,\xi_n)$ then $\Delta(f)$ has the symbol
$\hat{u}^n\frac{a(\xi_1,\ldots,\xi_n)}{1-(\xi_1+\cdots+\xi_n)^2}$.

All the definitions remain the same as in the local case with the amendment
$\ring\to\ring_{\Delta}$.

We consider a formal recursion operator for the equation
(\ref{CHe}). First of all we introduce a linear term to the equation
(\ref{CHe}) by the change of variable $m\mapsto m+1$ (note that
$\Delta(1)=1$) and consider the equation:
\begin{equation}
\label{CHe1} m_t=c\Delta(m_1)+cm\Delta(m_1)+\Delta(m)m_1+m_1:=F.
\end{equation}
Its symbolic representation reads:
\begin{equation*}
F\mapsto \hat{m}\omega(\xi_1)+\hat{m}^2a(\xi_1,\xi_2),
\end{equation*}
where $\omega(\xi_1)=\frac{c\xi_1}{1-\xi_1^2}+\xi_1,\quad
a(\xi_1,\xi_2)=\frac{c\xi_1+\xi_2}{2(1-\xi_1^2)}+\frac{c\xi_2+\xi_1}{2(1-\xi_2^2)}.$
The following statement holds \cite{mn2}:
\begin{The} The first two coefficients of the formal recursion
operator
\[
\Lambda=\eta+\hat{m}\phi_1(\xi_1,\eta)+\hat{m}^2\phi_2(\xi_1,\xi_2,\eta)+\cdots
\]
for the equation (\ref{CHe1}) are quasi-local if and only if $c=2$ or $c=3$.
\end{The}
The case $c=2$ corresponds to the Camassa-Holm equation, while the case $c=3$
corresponds to the Degasperis-Processi equation. In fact, one can show
\cite{HoneWang} that equation (\ref{CHe1}) (or equation (\ref{CHe})) with $c=2$ or
$c=3$ possesses an infinite dimensional algebra of {\it local} higher symmetries in
variable $m$ even if the equation is non-local.

\section{Integrable Boussinesq type equations}\label{sec_bouss}

In this section we give a brief account of our results (see details in
\cite{mnw07,nw07}) on integrable systems of the form:
\begin{equation}\label{bousssys}
\left\{\begin{array}{l} u_t=v_r,\\
v_t=\alpha u_{p-r}+\beta v_{q}+F(u,u_1,...,u_{p-r-1}, v,v_1, ..., v_{q-1}),
\end{array}\right.
\end{equation}
where $p>q\ge r\ge 0,\ \alpha,\beta\in\bbbc$. System (\ref{bousssys}) can be reduced
to a single second order (in time) non-evolutionary equation of order $p$:
\[ w_{tt} =\alpha  w_{p}+\beta  w_{q,\, t}+ F(w_r,w_{r+1},\ldots,
w_{p-1},w_t,w_{1,\, t},\ldots , w_{q-1,\, t} )\]
in the variable $w$, such that $v=\partial_t w,\ u= w_r$ and $w_k$ denotes
$\partial_x^k w$. If function $F$ does not depend on $v, v_1, \ldots , v_{r-1}$ then
one can eliminate $v$ from the second equation and rewrite the system in the form
\begin{equation}\label{eq1}
u_{tt}=\alpha u_p+\beta u_{t,q}+K(u,u_1,u_{2},...,u_{p-1},u_t,u_{t , 1},u_{t ,
2},...,u_{t, q-1})\, ,
\end{equation}
where $K=D_x^r(F)$.

The famous integrable Boussinesq equation \cite{zakharov}
\begin{equation}\label{bouss1}
u_{tt}=u_{xxxx}+(u^2)_{xx}
\end{equation}
belongs to this class.
Recently all integrable equations of the form
\begin{equation*}\label{shabclass}
u_{tt}=u_{xxx}+F(u,u_x,u_{xx},u_t,u_{t\, x})
\end{equation*}
have been classified and comprehensively studied in \cite{hss}. In particularly, it
has been shown that equation
\begin{equation*}\label{shabclass1}
w_{tt}=w_{xxx}+3w_xw_{t\, x}+w_{xx}w_t-3w_x^2w_{xx}.
\end{equation*}
is integrable.

Sixth order ($p=6$) integrable equations of the form (\ref{eq1}) can be obtained as
reductions of the Sato hierarchies corresponding to KP, BKP and CKP equations
\cite{satoKP,satoBKP} as well as derived from equations studied by Drinfeld and
Sokolov \cite{mr86h:58071,mr84m:35104}.

A non-evolutionary equation
\begin{equation}\label{eqnon}
u_{tt}=K(u,u_x,u_{xx},\cdots,\partial^n_x u,u_t,u_{t\, x},u_{t\,
xx},\cdots,\partial^m_x u_{t})\, ,
\end{equation}
can always be replaced by a system of two evolutionary equations
\begin{eqnarray*}
\left\{\begin{array}{l}\label{evsys0}
u_t=v,\\
v_t=K(u,u_x,u_{xx},...,\partial^n_x u,v,v_{x},v_{xx},...,\partial ^m_x v).
\end{array}\right .
\end{eqnarray*}

Non-evolutionary equation (\ref{eqnon}) may have other representations in the
evolutionary form. If $K=D_x (G)$, then the
system of evolutionary equations
\begin{equation*}\label{sys2}
u_t=v_x\, ,\qquad
v_t=G\, .
\end{equation*}
also represents (\ref{eqnon}).

For example, eliminating variable $v$ from the following systems

Case 1:
\begin{equation}\label{sysbouss1}
u_t=v\, ,\qquad
v_t=u_{xxxx}+(u^2)_{xx}\, ,
\end{equation}

Case 2:
\begin{equation}\label{sysbouss2}
u_t=v_x\, ,\qquad
v_t=u_{xxx}+(u^2)_{x}\, ,
\end{equation}

Case 3:
\begin{equation}\label{sysbouss3}
u_t=v_{xx}\, ,\qquad
v_t=u_{xx}+u^2\, ,
\end{equation}
we obtain the same Boussinesq equation (\ref{bouss1}) on variable
$u$.

There is a subtle, but important difference among the above representations of the
Boussinesq equation (\ref{bouss1}). The system (\ref{sysbouss1}) has only a finite
number of local infinitesimal symmetries (i.e. symmetries whose generators can be
expressed in terms of $u,v$ and a finite number of their derivatives) while equations
(\ref{sysbouss2}) and (\ref{sysbouss3}) have infinite hierarchies of local
symmetries. The reason is simple -- infinitesimal symmetries of equation
(\ref{sysbouss2}), which depend on the variable $v$, cannot be expressed in terms of
$u$ and its derivatives, since formally $v=D_x^{-1}u_t$.  Equation (\ref{sysbouss2})
possesses only a finite number of symmetries that do not depend explicitly on the
variable $v$.

Homogeneous equations play central role in the theory of integrable equations.
\begin{Pro}\label{prohombous}
Let system (\ref{bousssys}) be ${\bf w}$--homogeneous of weight $\mu$, then ${\bf
w}=(w_u, w_u+\mu-r)$ and
\begin{enumerate}
 \item [i.]  if $p$ is even, $p=2n,\ n\in\bbbn$, then $\mu=q=n$,
\item[ii.]  if $p$ is odd, $p=2n+1,\ n\in\bbbn$, then $\beta =0,\ \mu=p/2,\ q=n+1$.
\end{enumerate}
\end{Pro}

System (\ref{bousssys}) we will write in the vector form
\begin{equation}\label{bousr}
 \bu _t=\bF,\quad \bF=(v_r,f)^T.
\end{equation}
Suppose $\bG=(g,h)^T$ is a generator of a symmetry, then
\[ [\bF,\bG]=0 \ \Longleftrightarrow\ D_x^r(h)=D_\bF(g),\quad D_\bG(f)=D_\bF(h).   \]
It follows from the first equation that $D_\bF(g)$ belongs to the image of $D_x^r$
and thus the second component of the symmetry generator $\bG$ can be expressed as
$h=D_x^{-r}(D_\bF(g))$ and the symmetry is completely defined by its first component.
Substitution into the second equation yields
\begin{equation*}\label{symbous}
  D_\bF^2(g)-D_x^rD_\bG(f)=0\, .
\end{equation*}
For approximate symmetry of degree $n$ we  obviously get
\begin{equation*}\label{approxsymbous}
  D_\bF^2(g)-D_x^rD_\bG(f)=o(\cR^n)\, .
\end{equation*}

Using the above observation for systems (\ref{bousssys}) we can restrict the action
of the recursion operator to the first component of symmetries.

For example, if we represent the Boussinesq equation (\ref{bouss1})
in the form of evolutionary system (\ref{sysbouss2}), then
\[ {\bF}_*=  \left(\begin{array}{cc}
0&D_x\\ D_x^3+2uD_x+2u_1&0\end{array}\right) \] and it is easy to verify that a
pseudo-differential operator
\begin{equation*}
\Lambda=\left(\begin{array}{cc}
3v+2v_1 D_x^{-1}& 4D_x^{2}+2u+u_1D_x^{-1}\\
\Lambda_{21} & 3v+v_1 D_x^{-1}
\end{array}
\right),
\end{equation*}
where $\Lambda_{21}=4D_x^4+10uD_x^2+15u_1D_x+9u_2+4u^2+(2u_3+4uu_1)D_x^{-1}$,
satisfies equation (\ref{recopsys}) and therefore is a recursion operator. We can
restrict the action of the recursion operator on the first component. If $g_1$ is the
first component of a symmetry of the Boussinesq equation (\ref{sysbouss2}), then
\[ g_2=\Re(S)=\left( 3v+2v_1 D_x^{-1}+(4D_x^{2}+2u+u_1D_x^{-1})D_t D_x^{-1}\right) (g_1) \]
is the first component of the next symmetry in the hierarchy. We call $\Re$ a
restricted recursion operator.

A space-shift, generated by ${\bf u}_1=(u_1,v_1)^{\tr}$, is a symmetry of the
Boussinesq equation (\ref{sysbouss2}). Taking $u_1$ as a seed, we can construct an
infinite hierarchy $g_{3k+1}=\Re^k(u_1)$ of symmetries of weights $3k+1,
k=0,1,2,\ldots$. For example,
\[ g_4=4v_3+4v_1 u+4 vu_1=4D_x (v_2+vu)\, . \]
We see that $g_4$ is a total derivative and therefore $g_7=\Re(g_4)\in \ring$ is the
next symmetry in the hierarchy, etc. The Boussinesq equation itself is not a member
of this hierarchy. If we take a seed symmetry, corresponding to the time-translation
$(v_1,u_3+2uu_1)^{\tr}$ we obtain another infinite hierarchy of symmetries
$g_{3k+2}=\Re^k(v_1), k=0,1,2,\ldots$. The Boussinesq equation does not have
symmetries of weight $3k, k\in\mathbb N$. One can show that $g_{3k+1}$ and $g_{3k+2}$
are elements of the ring $\ring$ for any $k\in \mathbb N$ and therefore $\Re$
generates two infinite hierarchies of symmetries of the Boussinesq equation.
Moreover, all symmetries from the both hierarchies commute with each other.

In general, for system (\ref{bousr}) a recursion operator is completely determined by
its two entries $\Lambda_{11}$ and $\Lambda_{12}$ (the first row of $\Lambda$). The
restricted recursion operator $\Re$ for system (\ref{bousr}) can be represented as
\begin{equation*}\label{restrR}
\Re=\Lambda_{11}+\Lambda_{12}D_x^{-r} D_t \, .
\end{equation*}

In symbolic representation system (\ref{bousssys}) takes the form
\begin{eqnarray}\label{bousssyssym}
\left\{\begin{array}{l} \hu_t=\hv\zeta_1^r,\\
\begin{array}{l}\hv_t=\alpha \hu\xi_1^{p-r}+\beta \hv \zeta_1^q \\\quad+\sum_{k\ge
2}\sum_{i=0}^k \hu^i\hv^{k-i}a_{i,k-i}(\xi_1,..,\xi_i,\zeta_1,..,\zeta_{k-i})
.\end{array}
\end{array}\right.
\end{eqnarray} It follows from Proposition \ref{prohombous} that for ${\bf
w}$--homogeneous systems (\ref{bousssys}) $q=p/2$ (for even $p$) and coefficients
$a_{i,j}(\xi_1,\ldots,\xi_i,\zeta_1,\ldots,\zeta_{j})$ are homogeneous polynomials of
degree
\begin{equation*}\deg(a_{i,j})=
 p+(j-1)r-(i+j-1)w_u+(j-1)r-\frac{1}{2}jp,
\end{equation*}
in variables $\xi_1,\ldots,\xi_i,\zeta_1,\ldots,\zeta_{j}$, they are symmetric in
$\xi_1,\ldots,\xi_i$ and in $\zeta_1,\ldots,\zeta_{j}$. If $\deg(a_{i,j})$ is not a
non-negative integer, then $a_{i,j}=0$. We shall assume that the weight $w_u>0$ and
thus the sum in (\ref{bousssyssym}) is finite due to only a finite number of non-zero
coefficients $a_{ij}$.

\subsection{Even order equations}

In this section we study homogeneous (with a positive weight $w_u>0$) even order
($p=2n$) equations (\ref{bousssys}) assuming $r=1$.

In symbolic representation such equations take of the form
\begin{eqnarray}\label{symbbusseven}
\left\{\begin{array}{l}
\hu_t=\hv\zeta_1\\
\hv_t=\hu\omega_1(\xi_1)+\hv\omega_2(\zeta_1)+\hu^2a_{20}(\xi_1,\xi_2)+\hu \hv
a_{11}(\xi_1,\zeta_1)+\\\quad
+\hv^2a_{02}(\zeta_1,\zeta_2)+\hu^3a_{30}(\xi_1,\xi_2,\xi_3)+\cdots=\hat{f},
\end{array}\right.
\end{eqnarray}
where $\omega_1(\xi_1)=\alpha \xi_1^{2n-1}$ and $\omega_2(\zeta_1)=\beta \zeta_1^n$
and without loss of generality we shall represent $\alpha$ in the form
\[ \alpha=\frac{\mu^2-\beta^2}{4} \]
and use parameters $\mu,\beta$ instead of $\alpha,\beta$.

Symmetries of system (\ref{symbbusseven}) are determined by their first component,
which in symbolic representation can be written in the form
\begin{eqnarray*}
&&\hu_{\tau}=\hu\Omega_{1}(\xi_1)+\hv\Omega_{2}(\zeta_1)+\hu^2A_{20}(\xi_1,\xi_2)+
\hu\hv A_{11}(\xi_1,\zeta_1)+\nonumber\\
&&\quad
+\hv^2A_{02}(\zeta_1,\zeta_2)+\hu^3A_{30}(\xi_1,\xi_2,\xi_3)+\cdots,\label{symsymb}
\end{eqnarray*}
where $\Omega_{1}(\xi_1),\ \Omega_2(\zeta_1),\ A_{ij} $ are
polynomials. For fixed $\Omega_{1}(\xi_1),\ \Omega_2(\zeta_1)$ the
coefficients $A_{ij} $ can be found recursively (there is a
generalisation of Theorem \ref{theorsym} to the case of many
dependent variables \cite{nw07}).

The Fr\'echet derivative ${\bf F}_*$ of the system
(\ref{bousr})with $r=1$ in the symbolic representation
 has the form
\begin{equation*}\label{astar}
\hat{{\bf F}}_*=\left(\begin{array}{cc}
0& \eta\\
\hat{f}_{*,u}& \hat{f}_{*,v} \end{array}\right)
\end{equation*}
where
\begin{eqnarray*}
&&\hat{f}_{*,u}=\omega_1(\eta)+2\hu a_{20}(\xi_1,\eta)+\hv a_{11}(\eta,\zeta_1)
+3\hu^2a_{30}(\xi_1,\xi_2,\eta)+\cdots\\
&& \hat{f}_{*,v}=\omega_2(\eta)+\hu a_{11}(\xi_1,\eta)+2\hv
a_{02}(\zeta_1,\eta)+\hu^2a_{21}(\xi_1,\xi_2,\eta)+\cdots
\end{eqnarray*}
Formal recursion operator can be defined as a $2\times 2$ matrix $\hat{R}$ whose
entries are formal series
\begin{eqnarray*}
&&\hat{R}_{11}=\phi_{00}(\eta)+\hu\phi_{10}(\xi_1,\eta)+\frac{1}{2}\hv\phi_{01}(\zeta_1,\eta)+
\hu^2\phi_{20}(\xi_1,\xi_2,\eta)+\nonumber\\
&&\qquad +\hu\hv\phi_{11}(\xi_1,\zeta_1,\eta)+\hv^2\phi_{02}(\zeta_1,\zeta_2,\eta)+\cdots\label{rec1}\\
&&\hat{R}_{12}=\psi_{00}(\eta)+\frac{1}{2}\hu\psi_{10}(\xi_1,\eta)+\hv\psi_{01}(\zeta_1,\eta)+
\hu^2\psi_{20}(\xi_1,\xi_2,\eta)+\nonumber\\
&&\qquad
+\hu\hv\psi_{11}(\xi_1,\zeta_1,\eta)+\hv^2\psi_{02}(\zeta_1,\zeta_2,\eta)+\cdots\,
,\label{rec2}
\end{eqnarray*}
and
\begin{equation*}
\hat{R}_{21}=\eta^{-1}\circ(\hat{R}_{11,t}+\hat{R}_{12} \circ
\hat{f}_{*,u}),\quad
\hat{R}_{22}=\eta^{-1}\circ(\hat{R}_{12,t}+\hat{R}_{12} \circ
\hat{f}_{*,v}+\hat{R}_{11}\circ  \eta) .
\end{equation*}
 satisfying equation
\begin{equation*}
\label{receq} \hat{R}_t=[\hat{{\bf F}}_*,\hat{R}]
\end{equation*}
and all the coefficients
$\phi_{ij},\psi_{ij},\,i,j=0,1,2,\ldots$ of the formal series $\hat{R}_{11}$ and
$\hat{R}_{12}$ are local. Approximate recursion operator of degree $k$ can be viewed
as a truncation of $\hat{R}$ (terms with $\hat{u}^i\hat{v}^j, i+j>k$ are omitted or
ignored), so the existence of approximate recursion operator is a necessary condition
for the existence of a formal recursion operator and as well necessary condition for
the existence of an infinite hierarchy of symmetries of equation
(\ref{symbbusseven}). For fixed $\phi_{00}(\eta),\psi_{00}(\eta)$ the coefficients of
a formal recursion operator can be found recursively, similar to the case of one
dependent variable (Theorem \ref{prolambda}). The property of locality of the
coefficients imposes constraints on the coefficients of equation (\ref{symbbusseven})
and eventually leads to the isolation of integrable systems (see details in
\cite{mnw07}). The latter test had been applied in \cite{mnw07} to the problem of
classification of even order ${\bf w}$--homogeneous integrable systems of the form
(\ref{symbbusseven}). The results obtained can be summarised as follows:

\subsubsection*{The 4th order equations}

It is easy to see that a homogeneous system (\ref{bousssys}) with
$p=4,r=1$  is linear if $w_u> 3$. In the case $w_u=3$ the only
possibility is $F=\gamma u^2,\ \gamma\ne 0$ which leads to a
non-integrable equation for any choice of $\alpha,\beta$ and
$\gamma$. Thus the weight $w_u$ can be equal to $2$ or $1$.

{\bf Case 1:} {The case of $w_u=2$}

The most general nonlinear homogeneous system of equations (\ref{bousssys}) with
$p=4,r=1,w_u=2$ (correspondingly $w_v=3$) is of the form:
\begin{eqnarray}
\label{eq4h2}
\left\{\begin{array}{l}
u_t=v_1\\
v_t=\alpha u_3+\beta v_2+c_1uu_1+c_2uv,
\end{array}\right.
\end{eqnarray}
where $c_1,c_2$ are arbitrary constants and at least one of them is not zero.

Without loss of generality we need to consider the following three
types of system (\ref{eq4h2}):
\begin{eqnarray}
\label{eq4h2_1} &&\left\{\begin{array}{l}
u_t=v_1\\
v_t=\frac{\mu^2-\beta^2}{4} u_3+\beta v_2+c_1uu_1+c_2uv,\  \mu\notin \{0,\pm\beta\}
\end{array}\right.\\
\label{eq4h2_2} &&\left\{\begin{array}{l}
u_t=v_1\\
v_t=v_2+c_1uu_1+c_2uv
\end{array}\right.\\
\label{eq4h2_3} &&\left\{\begin{array}{l}
u_t=v_1\\
v_t=-\frac{1}{4}u_3+v_2+c_1uu_1+c_2uv
\end{array}\right.
\end{eqnarray}
The first system represents the generic case $\alpha\notin \{0, -\frac{\beta^2}
{4}\}$, while the other two represent the degenerate
cases.

\begin{The} \label{pro4w2} System (\ref{eq4h2_1}) possesses two formal recursion operators with
$\phi_{00}(\eta)=\eta,\,\psi_{00}(\eta)=0$
and $\phi_{00}(\eta)=0,\,\psi_{00}(\eta)=\eta$ if and only if $\beta=c_2=0$.
By re-scalings it can be put in the form
\begin{eqnarray}
\label{eq4h2a}
\left\{\begin{array}{l}
u_t=v_1\\
v_t=u_3+2uu_1
\end{array}\right.
\end{eqnarray}
Systems (\ref{eq4h2_2}) and (\ref{eq4h2_3}) are not integrable, they do not possess a formal
recursion operator with $\phi_{00}=0,\,\psi_{00}=\eta$ unless
$c_1=c_2=0$.
\end{The}
System (\ref{eq4h2a}) represents the Boussinesq equation (\ref{bouss1}), which is
known to be integrable. In the Proposition \ref{pro4w2} and below by re-scalings we
mean an invertible change of variable of the form:
\[
u\mapsto \alpha_1\,u,\quad  v\mapsto \alpha_2\,v,\quad x\mapsto \alpha_3\,x,\quad
t\mapsto \alpha_4\,t,\qquad \alpha_i\in\bbbc\, .\]

{\bf Case 2:} {The case of $w_u=1$}

The most general homogeneous system of equations (\ref{bousssys}) with
$p=4,r=1,w_u=1$ is of the form
\begin{eqnarray}
\label{eq4h1}
\left\{\begin{array}{l}
u_t=v_1\\
v_t=\alpha u_3+\beta
v_2+c_1uu_2+c_2u_1^2+c_3u_1v+c_4uv_1+c_5v^2\\
\qquad +c_6u^2u_1+c_7u^2v+c_8u^4
\end{array}\right.
\end{eqnarray}
where $c_i,\,i=1\ldots 8$ are arbitrary constants and we assume that
at least one of the coefficients $c_1,\ldots, c_5$ is not zero.
Without loss of generality we consider the following three types of
the system (\ref{eq4h1}):
\begin{eqnarray}
\label{eq4h1_1} \left\{\begin{array}{l}
u_t=v_1\\
v_t=\frac{\mu^2-\beta^2}{4} u_3+\beta
v_2+c_1uu_2+c_2u_1^2+c_3u_1v+c_4uv_1+c_5v^2\\
\qquad+c_6u^2u_1+c_7u^2v+c_8u^4,
\end{array}\right.
\end{eqnarray}
where $\mu\notin \{0,\pm\beta\},\,\,\,\mu,\beta\in {\bbbc}$ and
\begin{eqnarray}
\label{eq4h1_2} &&\left\{\begin{array}{l}
u_t=v_1\\
v_t=v_2+c_1uu_2+c_2u_1^2+c_3u_1v+c_4uv_1+c_5v^2\\
\qquad+c_6u^2u_1+c_7u^2v+c_8u^4
\end{array}\right.\\
\label{eq4h1_3} &&\left\{\begin{array}{l}
u_t=v_1\\
v_t=-\frac{1}{4}u_3+v_2+c_1uu_2+c_2u_1^2+c_3u_1v+c_4uv_1\\
\qquad+c_5v^2+c_6u^2u_1+c_7u^2v+c_8u^4
\end{array}\right.
\end{eqnarray}

\begin{The}
System (\ref{eq4h1_1}) possesses two  formal recursion operators
with $\phi_{00}(\eta)=\eta,\,\psi_{00}(\eta)=0$ and
$\phi_{00}(\eta)=0,\,\psi_{00}(\eta)=\eta$ if and only if (up to
re-scalings) it is one of the list
\begin{eqnarray}
&&\left\{\begin{array}{l}
u_t=v_1\\
v_t=u_3+u_1^2
\end{array}\right.
 \label{eq4h1a}\\
&&\left\{\begin{array}{l}
u_t=v_1\\
v_t=u_3+2u_1v+2u^2u_1
\end{array}\right.
\label{eq4h1b}\\
&&\left\{\begin{array}{l}
u_t=v_1\\
v_t=u_3+2u_1v+4uv_1-6u^2u_1
\end{array}\right.
\label{eq4h1c}\\
&&\left\{\begin{array}{l}
u_t=v_1\\
v_t=u_3+4uu_2+3u_1^2-v^2+6u^2u_1+u^4
\end{array}\right.
\label{eq4h1d}\\
&&\left\{\begin{array}{l}
u_t=v_1\\
v_t=\alpha u_3+v_2+4 \alpha u u_2+3 \alpha u_1^2+u_1 v +2 u v_1-v^2\\\qquad+6 \alpha
u^2 u_1+u^2 v+\alpha u^4,\,\,\alpha\ne -\frac{1}{4}
\end{array}\right. \label{eq4h1e}
\end{eqnarray}
System (\ref{eq4h1_2}) possesses a formal recursion operator  with
$\phi_{00}(\eta)=0$, $\psi_{00}(\eta)=\eta$ if and only if (up to re-scalings) it is
one of the list
\begin{eqnarray}
&&\left\{\begin{array}{l}
u_t=v_1\\
v_t=v_2+2 u v_1
\end{array}\right.
\label{eq4h1f}\\
&&\left\{\begin{array}{l}
u_t=v_1\\
v_t=v_2-u_1^2+2 u_1 v-v^2
\end{array}\right.
\label{eq4h1g}\\
&&\left\{\begin{array}{l}
u_t=v_1\\
v_t=v_2-2 u u_2-2 u_1^2+2 u_1 v+6 u v_1-12 u^2 u_1
\end{array}\right.\label{eq4h1h}
\end{eqnarray}
System (\ref{eq4h1_3}) possesses two formal recursion operators with
$\phi_{00}(\eta)=\eta$, $\psi_{00}(\eta)=0$ and $\phi_{00}(\eta)=0$,
$\psi_{00}(\eta)=\eta$  if and only if (up to re-scalings) it is
\begin{eqnarray}
\left\{\begin{array}{l}
u_t=v_1\\
v_t=-\frac{1}{4}u_3+v_2-u u_2-\frac{3}{4}u_1^2+u_1 v +2 u
v_1-v^2\\\qquad-\frac{3}{2}u^2 u_1+u^2 v-\frac{1}{4} u^4.
\end{array}\right. \label{eq4h1ep}
\end{eqnarray}

\end{The}
Equation (\ref{eq4h1a}) is a potential version of the Boussinesq equation. Equations
(\ref{eq4h1b}) and (\ref{eq4h1c}) are known to be integrable. Corresponding Lax
representations and references can be found in \cite{mr93g:35108}.

Equations (\ref{eq4h1d}) and  (\ref{eq4h1e})  can be mapped into linear equations
$$ w_{tt}=w_4\, ,\qquad w_{tt}=\alpha w_4+w_{2t} $$
respectively by the Cole-Hopf transformation $u=(\log w)_x$.

Equation (\ref{eq4h1f}) can be reduced to the Burgers equation with
time independent forcing
$$ u_t=u_2+2 uu_1+w_1 \, ,\qquad w_t=0$$
by invertible transformation $v=w+u_1+u^2$. The latter can be
linearised by the Cole-Hopf transformation.

Equation (\ref{eq4h1g}) can be reduced to the system
$$u_t=u_2+w_1\, ,\qquad w_t=-w^2$$
 by a simple invertible the change of the variable $v=w+u_1$. System (\ref{eq4h1g}) provides
an example of an equation that possesses neither higher symmetries
nor a recursion operator. However, a formal recursion operator does
exist and therefore it is in the list of the Proposition. Its
integration can be reduced to the integration of a linear
nonhomogeneous heat equation with a source term of a special form.

By a simple shift of the variable $v=w+u_1+2 u^2$,
system (\ref{eq4h1h}) can be transformed in the form
\begin{equation}\label{eq4h1hsys}
 u_t=u_2+4u u_1 +w_1\, ,\qquad w_t=2D_x (uw)\, .
\end{equation}
System (\ref{eq4h1h}) possess an infinite hierarchy of symmetries
of all orders generated by a recursion operator
\[ \Re=-2u+2 u_1 D_x^{-1}+D_t D_x^{-1}\, \]
starting from the seed $u_1$.

Equation (\ref{eq4h1ep}) is a particular case of (\ref{eq4h1e}) corresponding to the
exceptional case $\alpha=-\frac{1}{4}$.

\subsubsection*{The 6th order equations}
Homogeneous 6th order ($p=6$) equations (\ref{bousssys}) with non-zero
quadratic terms correspond to $w_u\le 5$. We restrict ourself with
the case $w_u>0$.  The weights $3,4,5$ do not lead to integrable
equations:

\begin{Pro} For weights $w_u=3, 4, 5$ there are no equations  possessing a
formal recursion operator.
\end{Pro}

{\bf Case 1:} {The case of $w_u=2$}

The most general homogeneous system (\ref{bousssys}) with $p=6,r=1,w_u=2$ can be written as
\begin{eqnarray}
&&\left\{\begin{array}{l}
u_t=v_1\\
v_t=\alpha u_5+\beta v_3+D_x(c_1uu_2+c_2u_1^2+c_5u^3)\\\qquad+c_3uv_1+c_4vu_1,
\end{array}\right.
 \label{eq6h2}
\end{eqnarray}
where $\alpha, \beta ,c_i,i=1,\ldots,4$ are arbitrary
constants and we assume that at least one of $c_1,\ldots, c_4$ is not zero.

Without loss of generality we consider the following three types of
the above system:
\begin{eqnarray}
&&\left\{\begin{array}{l}
u_t=v_1\\
v_t=\frac{\mu^2-\beta^2}{4} u_5+\beta
v_3+D_x(c_1uu_2+c_2u_1^2+c_5u^3)\\\qquad+c_3uv_1+c_4vu_1,\quad \mu\notin
\{0,\pm\beta\},\,\,\,\mu,\beta\in {\bbbc}
\end{array}\right.
 \label{eq6h2_1}\\
&&\left\{\begin{array}{l}
u_t=v_1\\
v_t=v_3+D_x(c_1uu_2+c_2u_1^2+c_5u^3)+c_3uv_1+c_4vu_1
\end{array}\right.
 \label{eq6h2_2}\\
&&\left\{\begin{array}{l}
u_t=v_1\\
v_t=-\frac{1}{4} u_5+v_3+D_x(c_1uu_2+c_2u_1^2+c_5u^3)\\\qquad+c_3uv_1+c_4vu_1
\end{array}\right.
 \label{eq6h2_3}
 \end{eqnarray}

\begin{The}\label{pro3}
If system (\ref{eq6h2_1}) possesses two formal recursion operators
with $\phi_{00}(\eta)=\eta,\,\psi_{00}(\eta)=0$ and
$\phi_{00}(\eta)=0,\,\psi_{00}(\eta)=\eta$ then, up to re-scalings,
it is one of the list
\begin{eqnarray}
&&\left\{\begin{array}{l}\label{eq6h2a}
u_t=v_1\\
v_t=2u_5+v_3+D_x(2uu_2+u_1^2+\frac{4}{27}u^3)
\end{array}\right.
\\
&&\left\{\begin{array}{l}\label{eq6h2b}
u_t=v_1\\
v_t=\frac{1}{5}u_5+v_3+D_x(uu_2+uv+\frac{1}{3}u^3)
\end{array}\right.
\\
&&\left\{\begin{array}{l} \label{eq6h2c}
u_t=v_1\\
v_t=\frac{1}{5}u_5+v_3+D_x(2uu_2+\frac{3}{2}u_1^2+2uv+\frac{4}{3}u^3)
\end{array}\right.
\end{eqnarray}
If system (\ref{eq6h2_2}) possesses a formal recursion operator with
$\phi_{00}(\eta)=0$, $\psi_{00}(\eta)=\eta$ then, up to re-scalings, it is one of the
list
\begin{eqnarray}
&&\left\{\begin{array}{l} \label{eq6h2d}
u_t=v_1\\
v_t=v_3+u v_1+u_1 v
\end{array}\right.
\\
&&\left\{\begin{array}{l} \label{eq6h2e}
u_t=v_1\\
v_t=v_3+2 u u_3+4 u_1 u_2-4 u_1 v-8 u v_1-24 u^2 u_1
\end{array}\right.
\end{eqnarray}
System (\ref{eq6h2_3}) does not possess a formal recursion operator for any
non-trivial choice of $c_i$.
\end{The}
Recursion operators and bi-Hamiltonian structure for these equations can be found in
\cite{mnw07}. Lax representations can be found in \cite{mr86h:58071,mr84m:35104,shabat05,mnw07,HNV1}.

{\bf Case 2:} {The case of $w_u=1$}

Homogeneous systems of equations (\ref{bousssys}) with $p=6,r=1,w_u=1$ can be written
in the form:
\begin{eqnarray}\label{eq6h1}
\left\{\begin{array}{l}
u_t=v_1\\
v_t=\alpha u_5+\beta v_3+c_1u_2^2+c_2u_1u_3+c_3uu_4+c_4u_2v+c_5u_1v_1\\
\quad+c_6uv_2
+c_7v^2+c_8u_1^3+c_9uu_1u_2 +c_{10}u^2u_3+c_{11}u^2v_1+\\
\quad+c_{12}uu_1v+c_{13} u^2u_1^2+ c_{14} u^3u_2+c_{15} u^3v+c_{16} u^4u_1+c_{17}
u^6,
\end{array}\right.
\end{eqnarray}
where $\alpha, \beta \in{\bbbc}$, all $c_i,i=1,\ldots,17$ are
arbitrary constants and at least one of $c_1,\ldots c_7$ is not
zero.

We need to consider the following cases of the system (\ref{eq6h1}):
\begin{eqnarray}\label{eq6h1_1}
\left\{\begin{array}{l}
u_t=v_1\\
v_t=\frac{\mu^2-\beta^2}{4} u_5+\beta v_3+c_1u_2^2+c_2u_1u_3+c_3uu_4
+c_4u_2v\\\quad+c_5u_1v_1+c_6uv_2+c_7v^2+c_8u_1^3++c_9uu_1u_2
+c_{10}u^2u_3\\
\quad+c_{11}u^2v_1 +c_{12}uu_1v +c_{13} u^2u_1^2+ c_{14} u^3u_2\\\quad+c_{15}
u^3v+c_{16} u^4u_1+c_{17} u^6
\end{array}\right.
\end{eqnarray}
where $\mu\notin \{0,\pm\beta\},\,\,\mu,\beta\in {\bbbc}$, and
\begin{eqnarray}
&&\left\{\begin{array}{l}
u_t=v_1\\
v_t=v_3+c_1u_2^2+c_2u_1u_3+c_3uu_4+c_4u_2v+c_5u_1v_1\\
\quad+c_6uv_2+c_7v^2+c_8u_1^3+c_9uu_1u_2 +c_{10}u^2u_3+c_{11}u^2v_1\\
\quad+c_{12}uu_1v+c_{13} u^2u_1^2+ c_{14} u^3u_2+c_{15} u^3v+c_{16} u^4u_1+c_{17} u^6
\end{array}\right.\label{eq6h1_2}\\
&& \left\{\begin{array}{l}
u_t=v_1\\
v_t=-\frac{1}{4}u_5+v_3+c_1u_2^2+c_2u_1u_3+c_3uu_4+c_4u_2v+c_5u_1v_1\\
\quad  +c_6uv_2+c_7v^2+c_8u_1^3+c_9uu_1u_2
+c_{10}u^2u_3+c_{11}u^2v_1\\\quad+c_{12}uu_1v+c_{13} u^2u_1^2+ c_{14} u^3u_2+c_{15}
u^3v+c_{16} u^4u_1+c_{17} u^6
\end{array}\right.\label{eq6h1_3}
\end{eqnarray}

\begin{The}
If system (\ref{eq6h1_1}) possesses two  formal recursion operators
with $\phi_{00}(\eta)=\eta,\,\psi_{00}(\eta)=0,$  and
$\phi_{00}(\eta)=0,\,\psi_{00}(\eta)=\eta,$ up to re-scalings, it is
one of the equations in the following list
\begin{eqnarray}
&&\left\lbrace \begin{array}{l}\label{eq6h1c} u_t=v_1\\
v_t=2u_5+v_3+u_2^2+2u_1u_3+\frac{4}{27}u_1^3\end{array}
\right. \\
&&\left\lbrace \begin{array}{l}
\label{eq6h1a} u_t=v_1\\
v_t=\frac{1}{5}u_5+v_3+u_1u_3+u_1v_1+\frac{1}{3}u_1^3\end{array}
\right. \\
&&\left\lbrace \begin{array}{l}\label{eq6h1b} u_t=v_1\\
v_t=\frac{1}{5}u_5+v_3+2u_1u_3+\frac{3}{2}u_2^2+2u_1v_1+\frac{4}{3}u_1^3\end{array}
\right.\\
&&\left\lbrace \begin{array}{l}\label{eq6h1e} u_t=v_1\\
v_t=\alpha u_5+v_3+10\alpha u_2^2+15\alpha u_1u_3+6\alpha
uu_4+vu_2\\\quad+3u_1v_1+3uv_2-v^2+ 15\alpha u_1^3+15\alpha u^2u_3+\\
\quad +60\alpha uu_1u_2+3uu_1v+3u^2v_1+45\alpha u^2u_1^2\\\quad+20\alpha
u^3u_2+u^3v+15\alpha u^4u_1+\alpha u^6,\,\,\alpha\ne-\frac{1}{4}\end{array}
\right. \\
&&\left\lbrace \begin{array}{l}\label{eq6h1f} u_t=v_1\\
v_t=u_5+6 u u_4+15 u_1 u_3+10 u_1^2-v^2+15 u^2 u_3+\\ \phantom{v_t=}+15 u_1^3 +60 u
u_1 u_2+45 u^2 u_1^2+20 u^3 u_2+15 u^4 u_1+u^6\end{array} \right.
\end{eqnarray}
If system (\ref{eq6h1_2}) possesses a  formal recursion operator
with $\phi_{00}(\eta)=0,\,\psi_{00}(\eta)=\eta,$ up to re-scalings,
it is one of the list
\begin{eqnarray}
&&\left\lbrace \begin{array}{l}\label{eq6h1d} u_t=v_1\\
v_t=v_3+u_1 v_1\end{array}
\right. \\
&&\left\lbrace \begin{array}{l}\label{eq6h1g} u_t=v_1\\
v_t=v_3+3 u_1 v_1+3 u v_2+3 u^2 v_1\end{array}
\right. \\
&&\left\lbrace \begin{array}{l}\label{eq6h1h} u_t=v_1\\
v_t=v_3-u_2^2+2 u_2 v-v^2\end{array}\right.
\end{eqnarray}
If system (\ref{eq6h1_3}) possesses two  formal recursion operators
with $\phi_{00}(\eta)=\eta,\,\psi_{00}(\eta)=0,$  and
$\phi_{00}(\eta)=0,\,\psi_{00}(\eta)=\eta,$ up to re-scalings, then
it is
\begin{eqnarray}
\left\lbrace \begin{array}{l}\label{eq6h1ep} u_t=v_1\\
v_t=-\frac{1}{4} u_5+v_3-\frac{5}{2} u_2^2-\frac{15}{4} u_1u_3-\frac{3}{2}
uu_4+vu_2\\\quad+3u_1v_1+3uv_2-v^2- \frac{15}{4} u_1^3-\frac{15}{4}
u^2u_3-15uu_1u_2+3uu_1v\\\quad+3u^2v_1-\frac{45}{4}
u^2u_1^2-5u^3u_2+u^3v-\frac{15}{4}u^4u_1-\frac{1}{4} u^6,\end{array} \right.
\end{eqnarray}
\end{The}
Recursion operators, bi-Hamiltonian structure and Lax representations of equations
(\ref{eq6h1c}), (\ref{eq6h1a}), (\ref{eq6h1b}) and (\ref{eq6h1d}) are discussed in
details in \cite{mnw07}.

Equations (\ref{eq6h1e}),(\ref{eq6h1f}),(\ref{eq6h1g}) and (\ref{eq6h1ep}) can be
linearised by a Cole-Hopf type transformation \cite{mnw07}.

Equation (\ref{eq6h1h}) has similar property as system (\ref{eq4h1g}).
It can also be reduced to a triangular system
\[u_t=u_3+w_1,\qquad w_t=-w^2\]
in  variables $u$ and $w=v-u_2$.

\subsubsection*{10th order equations}
In this section we present three examples of 10th order integrable
non-evolutionary equations.

\begin{Pro}
The following systems possess infinite hierarchies of higher
symmetries:
\begin{eqnarray}
&&\left\{ \begin{array}{ll}
u_t=v_1\\
v_t=\frac{9}{64}u_9+v_5+D_x\left( 3uu_6+9u_1u_5+\frac{65}{4}u_2u_4
+\frac{35}{4}u_3^2+\right. \\
\quad  +2u_1v_1+4uv_2+20u^2u_4+80uu_1u_3+60uu_2^2+88u_1^2u_2\\\quad
\left.+\frac{256}{5}u^3u_2+\frac{384}{5}u^2u_1^2 +\frac{1024}{125}u^5\right)
\end{array} \right.\label{ex5}\\
&&\left\{\begin{array}{l}
u_t=v_1\\
v_t=-\frac{1}{54}u_{9}+v_5+\frac{5}{6}u_7u_1
+\frac{5}{3}u_6u_2+\frac{5}{2}u_5u_3+\frac{25}{12}u_4^2
\\
\quad-5u_3v_1-\frac{15}{2}u_2v_2-10u_1v_3 -\frac{45}{4}u_5u_1^2-\frac{75}{2}u_1u_2u_4
\\\quad-\frac{75}{4}u_3^2u_1 -\frac{75}{4}u_2^2u_3+\frac{45}{2}u_1^2 v_1+
\frac{225}{4}u_3u_1^3 +\frac{675}{8}u_2^2u_1^2-\frac{405}{16}u_1^5
\end{array}\right.\label{ex4b}\\
&&\left\{\begin{array}{l}
u_t=v_1\\
v_t=v_5+2 u_2 u_5+6 u_3 u_4-6
 u_3 v-22 u_2 v_1-30 u_1 v_2
\\ \quad -20 u v_3
+96 u u_1 v +96 u^2 v_1+120 D_x (4 u^3 u_2 +6 u^2 u_1^2)\\\quad -2 D_x ( 8 u^2 u_4+32
u u_1 u_3 +13 u_1^2 u_2 +24 u u_2^2)  -3840 u^4 u_1
\end{array}\right.\label{ex4c}
\end{eqnarray}
\end{Pro}
Bi-Hamiltonian structures and recursion operators for equations (\ref{ex5}),
(\ref{ex4b}) and (\ref{ex4c}) as well as the Lax representation for equation
(\ref{ex5}) can be found in \cite{mnw07}. System (\ref{ex4c}) has been considered
independently by authors of \cite{HNV1, HNV2} and \cite{Serg}, where the
corresponding Lax representation has been obtained. The Lax representations for
equation (\ref{ex4b}) is still not known.

\subsection{Odd order equations}\label{oddorderbousseq}
In this section we formulate the diagonalisation method and globally
classify homogeneous (with a positive weight $w_u>0$) odd order
($p=2n+1$) equations (\ref{bousssys}) in the same spirit as we did
for the scalar homogeneous evolution equations in Section
\ref{sec3}. For the details, we refer the reader to the recent paper
\cite{nw07}.

A symmetry of an odd order homogeneous equation (\ref{bousssys})
always starts with linear terms. A homogeneous symmetry in the
symbolic representation starts either with $u\xi_1^m$ or with
$v\zeta_1^{m+r}$. Without loss of generality we have
\begin{equation}
\label{sym1} u_{\tau}=G=u\xi_1^m+\sum_{s\ge 2}\sum_{j=0}^s
u^jv^{s-j}A_{j,s-j}(\xi_1,..,\xi_j,\zeta_1,..,\zeta_{s-j}),\quad m>1
\end{equation}
or
\begin{equation}
\label{sym2} u_{\tau}=G=v\zeta_1^{m+r}+\sum_{s\ge 2}\sum_{j=0}^s
u^jv^{s-j}A_{j,s-j}(\xi_1,..,\xi_j,\zeta_1,..,\zeta_{s-j}),\quad m>0
\end{equation}
Functions $A_{i,j}(\xi_1,\ldots,\xi_i,\zeta_1,\ldots,\zeta_j)$ in
(\ref{sym1}) and (\ref{sym2}) are homogeneous polynomials in their
variables, symmetric with respect to arguments $\xi_1,\ldots,\xi_i$
and $\zeta_1,\ldots,\zeta_j$. These functions can be explicitly
determined in the terms of system (\ref{bousssyssym}), the symbolic
representation of system (\ref{bousssys}), from the compatibility
conditions.

Let us first concentrate on how to compute the Lie bracket between the linear part of
system (\ref{bousssyssym}) denoted by $K^1$, i.e.
\begin{eqnarray*}
K^1=\left(\begin{array}{c}v\zeta_1^r\\u\xi_1^{2n+1-r}\end{array}\right)
=L\left(\begin{array}{c}u\\v\end{array}\right), \quad  L=\left(\begin{array}{cc} 0&
\eta^r\\ \eta^{2n+1-r}& 0\end{array}\right)
\end{eqnarray*}
and any pair of differential polynomials. We know its symbolic representation takes
simple and elegant form if matrix $L$ is diagonal \cite{MR2070382}. Inspired by this,
we shall diagonalise matrix $L$, produce the required formula in new variables and
then transform back to the original variables.

Notice that matrix $L$ has two eigenvalues
$\pm\eta^{n+\frac{1}{2}}$. Therefore, there exists a linear
transformation
\[
T=\left(\begin{array}{cc}1&1\\ \eta^{n+\frac{1}{2}-r}&
-\eta^{n+\frac{1}{2}-r}\end{array}\right)\]
such that
\[
T^{-1}LT=\diag(\eta^{n+\frac{1}{2}},-\eta^{n+\frac{1}{2}}).
\]
Let us introduce new variables $\hu$ and $\hv$
\[
\left(\begin{array}{c}u\\v\end{array}\right)=T\left(\begin{array}{c}\hu\\\hv\end{array}
\right)=\left(\begin{array}{c}
\hu+\hv\\
\hu\xi_1^{n+\frac{1}{2}-r}-\hv\zeta_1^{n+\frac{1}{2}-r}\end{array}\right).
\]
Equally, we have
\[
\left(\begin{array}{c}\hu\\\hv\end{array}\right)=T^{-1}\left(\begin{array}{c}u\\v
\end{array}\right)=\frac{1}{2}
\left(\begin{array}{c}u+\eta^{-n-\frac{1}{2}+r}(v)\\
u-\eta^{-n-\frac{1}{2}+r}(v)\end{array}\right).
\]
The new variables $\hu$ and $\hv$ have the same weight, i.e., $w_{\hu}=w_{\hv}=w_u$.
Without causing a confusion we assign the same symbols $\xi$ and $\zeta$ for the
symbolic representation of the ring generated by $\hu$, $\hv$ and their derivatives.
The exponents of symbols can be half-integer, which corresponds to
half-differentiation in $x$-space.

\begin{Pro}
In variables $\hat{u}$ and $\hat{v}$, system (\ref{bousssyssym}) takes the form
\begin{eqnarray*}
\left\{\begin{array}{l}\hu_t=\hu\xi_1^{n+\frac{1}{2}}+\frac{1}{2 g}\sum_{s\ge
2}\sum_{l=0}^s\hu^l\hv^{s-l}\hat{a}_{l,s-l}(\xi_1,..,\xi_l,\zeta_1,
..,\zeta_{s-l})\\
\hv_t=-\hv\zeta_1^{n+\frac{1}{2}}-\frac{1}{2g}\sum_{s\ge
2}\sum_{l=0}^s\hu^l\hv^{s-l}\hat{a}_{l,s-l}(\xi_1,..,\xi_l,\zeta_1, ..,\zeta_{s-l})
\end{array}\right. ,
\end{eqnarray*}
where $g=(\xi_1+\cdots+ \xi_l+\zeta_1+\cdots+\zeta_{s-l})^{n+\frac{1}{2}-r}$ and
$\hat{a}_{l,s-l}$ are defined in terms of
${a}_{i,s-i}(\xi_1,\ldots,\xi_i,\zeta_1,\ldots,\zeta_{s-i}),\  i=0,\ldots,s$ as
follows:
\begin{eqnarray*}
&&\hat{a}_{l,s-l}(\xi_1,..,\xi_l,\zeta_1,..,\zeta_{s-l})=
\sum_{i=0}^s\sum_{p=max\{0,l-s+i\}}^{min\{i,l\}} C_i^pC_{s-i}^{l-p}\\
&&(-1)^{s-i-l+p}\langle\langle
a_{i,s-i}(\xi_1,..,\xi_p,\zeta_1,..,\zeta_{i-p},\xi_{p+1},
\ldots,\xi_l,\zeta_{i-p+1},..,\zeta_{s-l})\\
&&(\xi_{p+1}\cdots\xi_j\zeta_{i-p+1}\cdots\zeta_{s-l})^{n+\frac{1}{2}-r}\rangle_{\cS^{\xi}_l}
\rangle_{\cS^{\zeta}_{s-l}},
\end{eqnarray*}
where $C_i^j$ are binomial coefficients defined by $ C_i^j=\frac{i!}{j! (i-j)!}. $
\end{Pro}

In variables $\hu$ and $\hv$, the linear parts of symmetries (\ref{sym1}) and
(\ref{sym2}) of system (\ref{bousssys}) are also diagonal matrices. We can now derive
the symmetry conditions for the transformed forms as in Theorem \ref{theorsym} for
the scalar case.  This leads to the explicit recursive relations between symmetry
(\ref{sym1}) or (\ref{sym2}) and system (\ref{bousssys}).

With explicit formulas at hand, we can prove the following theorem,
crucial for global classification:

\begin{The}
Assume homogeneous system (\ref{bousssys}) ($p=2n+1$) with $w_u>0$ possesses a
symmetry. Suppose there is another system of the same weight and of the same form
\begin{eqnarray}
\left\{\begin{array}{l}\label{eqb} u_t=v\zeta_1^r,\\
v_t=u\xi_1^{2n+1-r}+\sum_{k\ge 2}\sum_{i=0}^k
u^iv^{k-i}b_{i,k-i}(\xi_1,..,\xi_k,\zeta_1,..,\zeta_{k-i}),
\end{array}\right.
\end{eqnarray}
whose quadratic terms equal to those of (\ref{bousssys}), that is,
$b_{i,2-i}(x,y)=a_{i,2-i}(x,y),\,i=0,1,2$. Then if system (\ref{eqb}) possesses a
symmetry of the same order, then equation (\ref{eqb}) and (\ref{bousssys}) are equal
and sharing the same symmetry.
\end{The}

This can be viewed as another version of Theorem \ref{MainR} for the
systems case: the existence of infinitely many approximate
symmetries of low degree together with one symmetry implies
integrability.

Now we formulate the classification theorem:

\begin{The}\label{Th2}
If a homogeneous system (\ref{bousssys}) with a positive weight $w_u>0$ of odd order
($p=2n+1$) possesses a hierarchy  of infinitely many higher symmetries, then it is
one of the systems in the following list up to re-scaling $u\mapsto\alpha
u,v\mapsto\beta v,t\mapsto\gamma t,x\mapsto\delta x$, where
$\alpha,\beta,\gamma,\delta\in\bbbc$:
\begin{eqnarray*}
&&\left\{\begin{array}{l}\nonumber u_t=v_1,\\
v_t=u_2+3uv_1+vu_1-3u^2u_1,
\end{array}\right.\\
&&\left\{\begin{array}{l}\label{Burg} u_t=v_1,\\
v_t=(D_x+u)^{2n}(u)-v^2,\quad n=1,2,3,\ldots .
\end{array}\right.
\end{eqnarray*}
\end{The}
These systems can be rewritten in the form of non-evolutionary
equations if we introduce a new variable $u=w_x$:
\[
w_{tt}=w_{xxx}+3w_xw_{t,x}+w_{xx}w_t-3w_x^2w_{xx}.
\]
\[
w_{tt}=(\partial_x+w_x)^{2n}(w_x)-w_t^2
\]
The first equation is known to be integrable \cite{hss}. The second
equation can be brought into linear $f_{tt}=\partial_x^{2n+1} f$ by
the Cole-Hopf transformation $w=\log(f)$. Its symmetries  are given
by
\[
u_{\tau_m}=D_x(D_x+u)^{m-1}u,\quad \,m=2,3,\ldots
\]
and
\[
u_{\tau_m}=D_x(v+D_t)(D_x+u)^{m-1}u.
\]
These symmetries correspond to the symmetries $f_{\tau_m}=f_m$ and
$f_{\tau_m}=f_{t,m}$ of equation $f_{tt}=f_{2n+1}$.

\section{Symmetry structure of $(2+1)$--dimensional integrable equations}\label{2+1}
This section is devoted to the study of $(2+1)$--dimensional
integrable equations. A famous example is the Kadomtsev-Petviashvili
(KP) equation
$$ u_t=u_{xxx}+6 u u_x +3 D_x^{-1} u_{yy}. $$
One of the main obstacles to extend the spectacular classification
results of $(1+1)$-dimensional integrable equations to the
$(2+1)$-dimensional case is that the equations themselves, their
higher symmetries and conservation laws are non-local, i.e. they
contain integral operators $D_x^{-1}$ or $D_y^{-1}$. In 1998,
Mikhailov and Yamilov, \cite{mr1643816}, introduced the concept of
quasi-local functions based on the observation that the operators
$D_x^{-1}$ and $D_y^{-1}$ never appear alone but always in pairs
like $D_x^{-1}D_y$ and $D_y^{-1}D_x$ for all known integrable
equations and their hierarchies of symmetries and conservation laws,
what enabled them to extend the symmetry approach of testing
integrability \cite{mr93b:58070} to the $(2+1)$-dimensional case.

The results in this section are based on a recent paper \cite{wang21}. We develop the
symbolic representation method to derive the hierarchies of $(2+1)$-dimensional
integrable equations from the scalar Lax operators and to study their properties
globally.  We prove that these hierarchies are indeed quasi-local as conjectured by
Mikhailov and Yamilov in 1998, \cite{mr1643816}.

\subsection{Quasi-local polynomials and Symbolic representation}
The basic definitions and notations of the ring of (commutative)
differential polynomials are similar to in section \ref{diffpoly1}.
The derivatives of dependent variable $u$ with respect to its
independent variables $x$ and $y$ are denoted by
$u_{ij}=\partial_x^i \partial_y^j u$. For small values of $i$ and
$j$, we sometimes write the indices out explicitly, that is
$u_{xxy}$ and $u$ instead of $u_{21}$ and $u_{00}$.

A differential monomial takes the form $u_{i_1 j_1}u_{i_2 j_2}\cdots u_{i_n j_n}.$ We
call $n$ the degree of the monomial. Let $\cR^n$ denote the set of differential
polynomials of degree $n$. The ring of differential polynomials is denoted by
$\cR=\oplus_{n\ge 1}\cR^n$. It is a differential ring with total $x$-derivation and
$y$-derivation
$$
D_x=\sum_{i, j\ge 0}u_{i+1,j}\frac{\partial} {\partial u_{ij}} \qquad \mbox{and}\quad
D_y=\sum_{i, j\ge 0}u_{i,j+1}\frac{\partial} {\partial u_{ij}}.
$$
Let us denote
\begin{eqnarray*}\label{Theta}
\Theta=D_x^{-1}D_y, \qquad \Theta^{-1}=D_y^{-1}D_x.
\end{eqnarray*}
The concept of {\bf quasi-local (commutative) polynomials $\cR_{\Theta}$} was
introduced in \cite{mr1643816}. Its definition is similar to the ring extension of
$\ring_H$ as in section \ref{bo-sec}. We consider a sequence of extensions of $\cR$
as follows:
\begin{equation*}
{\cal R}_{\Theta^0}=\ring\, , \quad {\cal R}_{\Theta^{n+1}}=\overline{{\cal
R}_{\Theta^n}\bigcup \Theta({\cal R}_{\Theta^n})\bigcup \Theta^{-1}({\cal
R}_{\Theta^n})}\, ,
\end{equation*}
where the sets $$\Theta({\cal R}_{\Theta^n})=\{ \Theta(f); f\in{\cal R}_{\Theta^n}\},
\quad \Theta^{-1}({\cal R}_{\Theta^n})=\{ \Theta^{-1}(f); f\in{\cal R}_{\Theta^n}\}$$
and the horizontal line denotes the ring closure. Each ${\cal R}_{\Theta^n}$ is a
ring and the index $n$ indicates the nesting depth of the operators $\Theta^{\pm1}$.
Clearly, we have
\[
{\cal R}_{\Theta^0}\subset{\cal R}_{\Theta^1}\subset{\cal
R}_{\Theta^2}\subset\cdots\subset{\cal R}_{\Theta^n}\subset\cdots\subset {\cal
R}_{\Theta^\infty}={\cal R}_\Theta.
\]

To define the symbolic representation of $\ring$, we replace
$u_{ij}$ by $\hu \xi^i \eta^j$, where $\xi$ and $\eta$ are symbols
(comparing to the definitions in section \ref{symbol}).
\begin{Def}
The symbolic representation of a differential monomial is defined as
$$u_{i_1,j_1}u_{i_2,j_2}\cdots u_{i_n,j_n}\longmapsto
\frac{\hu^n}{n!} \sum_{\sigma\in \cS_n} \xi^{i_1}_{\sigma(1)}
\eta^{j_1}_{\sigma(1)}\cdots \xi^{i_n}_{\sigma(n)}\eta^{j_n}_{\sigma(n)}\ .$$
\end{Def}
The result of action of operators $\Theta^{\pm 1}$ on a monomial $$\hu^k
a(\xi_1,\ldots,\xi_k,\eta_1,\ldots,\eta_k)$$ in the symbolic representation is given
by  $$\hu^k a(\xi_1,\ldots,\xi_k,\eta_1,\ldots,\eta_k)\left(\frac{\eta_1+\cdots
+\eta_k}{\xi_1+\cdots +\xi_k}\right)^{\pm 1}.$$ By induction, we can define the
symbolic representation of any element in $\cR_\Theta$, which is a rational function
with its denominator being the products of the linear factors. The expression of the
denominator uniquely determines how the operator $\Theta^{\pm1}$ is nested. For
example, the symbol of $u_{x}(\Theta u) \Theta^{-1} u\in \cR_{\Theta^1}$ is
\begin{eqnarray*}
&&\frac{\hu^3}{3} (\xi_1 \frac{\eta_2} {\xi_2}\frac{\xi_3}{\eta_3} + \xi_2
\frac{\eta_1} {\xi_1}\frac{\xi_3}{\eta_3}+\xi_3 \frac{\eta_2}
{\xi_2}\frac{\xi_1}{\eta_1})\ .
\end{eqnarray*}

The symbolic representations of pseudo-differential operators in $(2+1)$--dimensional
case are similar to for the case of one spatial variable in section \cite{wang21}.
However, we assign a special symbol $X$ to the operator $D_x$ in contrast to the
symbol $Y$ for the operator $D_y$.

The extension of the symbolic representation from one dependent
variable to several dependent variables is straightforward. We need
to assign new symbols for each of them such as assigning $\xi^{(1)},
\eta^{(1)}$ for $u$ and $\xi^{(2)}, \eta^{(2)}$ for $v$ and so on.

\subsection{Lax formulation of $(2+1)$--dimensional integrable equations}\label{Sec3}
We give a short description of construction of $(2+1)$-dimensional integrable
equations from a given scalar Lax operator based on the well-known Sato approach. For
details on the Sato approach for the $(1+1)$-dimensional case, see the recent books
\cite{BS01,kp00} and related references in them.

Let $H$ be an $m$-th order pseudo-differential operator in two
spatial variables of the form
$$
H=-D_y+a_m D_x^m+a_{m-1} D_x^{m-1}+\cdots +a_0 +a_{-1} D_x^{-1}+\cdots, \quad m\geq
0,
$$
where coefficients $a_k$ are functions of $x, \ y$. Let the commutator be the bracket
on the set of pseudo-differential operators. Thus, the set of pseudo-differential
operators forms a Lie algebra. For an integer $k<m$, we split into
\begin{eqnarray*}
&&H_{\geq k}=a_m D_x^m+a_{m-1} D_x^{m-1}+\cdots +a_k D_x^{k}\\
&&H_{<k}=H-H_{\geq k}=-D_y+a_{k-1} D_x^{k-1}+\cdots
\end{eqnarray*}
This operator algebra decomposes as a direct sum of two subalgebras in both
commutative and noncommutative cases when $k\in \{0,1\} $. Similar to the
$(1+1)$-dimensional case, such decompositions are naturally related with
integrability and lead to admissible scalar Lax operators for the case of $(2+1)$
dimensions:
\begin{enumerate}
\item[{\bf a}.] $ \quad k=0: \quad n\geq 2, $

$L=D_x^n+u^{(n-2)}D_x^{n-2} +u^{(n-3)}D_x^{n-3}+\cdots +u^{(0)} -D_y $;
\item[{\bf b}.] $ \quad k=1: \quad n\geq 2,$

$L=D_x^n+u^{(n-1)}D_x^{n-1} +u^{(n-2)}D_x^{n-2}+\cdots +u^{(0)} +D_x^{-1}
u^{(-1)}-D_y $;
\item[{\bf c}.] $ \quad k=1: \quad L=u^{(0)} +D_x^{-1} u^{(-1)}-D_y$;
\end{enumerate}
where $u^{(i)}$ are functions of two spatial variables $x,\ y$. We often use $u,v, w,
\cdots$ in the examples. For the KP equation, the Lax operator is the case {\bf a}
when $n=2$, namely, $L=D_x^2+u-D_y$.

Let $S= D_x+a_0+a_{-1}D_x^{-1}+\cdots$. For any operator $L$ listed
in cases {\bf a}, {\bf b} and {\bf c}, the relation
\begin{eqnarray}
&&[S,\ L]:= S L-L S =0,\label{S}
\end{eqnarray}
uniquely determines the operator $S$ by taking the integration
constants to be zeros. Furthermore, we have $[S^n,\ L]=0$ for any
$n\in \bbbn$. For each choice of $i$, we introduce a different time
variable $t_i$ and define the Lax equation by
\begin{eqnarray}\label{Laxeq}
\frac{\partial L}{\partial {t_i}}=[{S^i}_{\geq k}, L],
\end{eqnarray}
where $k$ is determined by the operator $L$ as listed in cases {\bf a}, {\bf b} and
{\bf c}.

\begin{The}\label{Th1}
For the operator $S$ uniquely determined by (\ref{S}), the flows defined by Lax
equations (\ref{Laxeq}) commute, i.e., $\partial_{ t_j} \partial_{t_i} L=
\partial_{ t_i} \partial_{t_j} L$.
\end{The}

\subsection{Lax formulation in symbolic representation}\label{Sec4}
We put the formalism of section \ref{Sec3} into the symbolic form.
The strategy is to do the calculation as much as possible without
symmetrisation and only perform the symmetrisation at the last stage
to get the uniqueness of the symbolic representation since the
symmetrisation complicates the calculation dramatically.

Let us assign the symbols $\xi^{(i)}, \eta^{(i)}$ for dependent variable $u^{(i)}$.
The symbolic representations of the admissible scalar Lax operators are
\begin{enumerate}
\item[{\bf a}.] $\quad k=0:\quad n\geq 2,$

$L=X^n-Y+{\hat u^{(n-2)}}X^{n-2} +\hu^{(n-3)}X^{n-3} +\cdots +\hu^{(0)}\  ;$
\item[{\bf b}.] $\quad  k=1:\quad n\geq 2,$

$\hat L=X^n-Y+\hu^{(n-1)}X^{n-1} +\cdots +\hu^{(0)} +
\hu^{(-1)}\frac{1}{X+\xi^{(-1)}}\   $;
\item[{\bf c}.] $ \quad k=1: \quad
\hat L=-Y+\hu^{(0)}+\hu^{(-1)}\frac{1}{X+\xi^{(-1)}}$;
\end{enumerate}
Here we only treat the case {\bf a}. The study of the cases {\bf b} and {\bf c} can
be found in \cite{wang21}.

It is convenient to consider formal series in the form
\begin{eqnarray}
&&S=X+\sum_{i=0}^{n-2} \hu^{(i)} a_1^{(i)}(\xi_1^{(i)},\eta_1^{(i)},X)\label{SA}\\
&+&\sum_{i_1=0}^{n-2} \sum_{i_2=0}^{n-2} \hu^{(i_1)} \hu^{(i_2)}a_2^{(i_1
i_2)}(\xi_{j_1}^{(i_1)},\eta_{j_1}^{(i_1)}, \xi_{j_2}^{(i_2)},\eta_{j_2}^{(i_2)},X)
+\cdots \, ,  \nonumber
\end{eqnarray}
where $n\geq 2$ and $a_i$ are functions of their specific arguments, the superindex
$i_s\in \{0,1,2,\cdots, n-2\}$ and the subindex $j_k$ is defined by the number of
$i_k$ in the list of $[i_1, i_2, \cdots, i_l]$. This implies that $j_1=1$ and
$j_k\geq 1$, $k=1,2,\cdots $. For example, when $i_1=i_2$, the arguments of function
$a_2^{i_1i_1}$ are $\xi_{1}^{(i_1)}$, $\eta_{1}^{(i_1)}$, $\xi_{2}^{(i_1)}$,
$\eta_{2}^{(i_1)}$ and $X$.

It is easy to check that
\begin{eqnarray*}
&&[X^n-Y,\ \phi(\xi_{j_1}^{(i_1)},\eta_{j_1}^{(i_1)},
\cdots,\xi_{j_l}^{(i_l)},\eta_{j_l}^{(i_l)},X)]=\nonumber\\
&=&N_l(\xi_{j_1}^{(i_1)},\eta_{j_1}^{(i_1)},
\cdots,\xi_{j_l}^{(i_l)},\eta_{j_l}^{(i_l)},X) \phi, \label{linear}
\end{eqnarray*}
where the polynomial $N_l$ is defined by
\begin{eqnarray*}
 N_l(\xi_1,\eta_1,\xi_2,\eta_2,\ldots, \xi_l,\eta_l;X)=(\sum_{i=1}^l \xi_i +X)^n-X^n
 -\sum_{i=1}^l \eta_i \ .\label{dem}
\end{eqnarray*}

\begin{Pro}\label{Pro1}
For any operator $L$ in case {\bf a}, if the formal series (\ref{SA}) satisfies the
relation $[S,\ L]=0$ (cf. (\ref{S})), we have for $l\geq 1$,
\begin{eqnarray*}
&& a_{l}^{(i_1i_2\cdots i_{l})}=
a_l(\xi_{j_1}^{(i_1)},\eta_{j_1}^{(i_1)},\cdots,\xi_{j_l}^{(i_l)},\eta_{j_l}^{(i_l)},X)\nonumber\\
&=& \prod_{r=1}^{l}(X+\sum_{s=r+1}^{l}\xi_{j_s}^{(i_s)})^{i_r}
b_{l}(\xi_{j_1}^{(i_1)},\eta_{j_1}^{(i_1)},..,\xi_{j_l}^{(i_l)},
\eta_{j_l}^{(i_l)},X) \label{Al}.
\end{eqnarray*}
The function $b_l$, $l\geq1$, is defined by
\begin{eqnarray*}
&&b_l(\xi_{j_1}^{(i_1)},\eta_{j_1}^{(i_1)},\cdots,\xi_{j_l}^{(i_l)},\eta_{j_l}^{(i_l)},X)=\nonumber\\
&=&\frac{c_l(\xi_{j_1}^{(i_1)},\eta_{j_1}^{(i_1)},\cdots,\xi_{j_l}^{(i_l)},\eta_{j_l}^{(i_l)},X)}
{N_l(\xi_{j_1}^{(i_1)},\eta_{j_1}^{(i_1)},\cdots,\xi_{j_l}^{(i_l)},\eta_{j_l}^{(i_l)},X)}
\label{Bf}
\end{eqnarray*}
with
\begin{eqnarray*}
&&c_l(\xi_{j_1}^{(i_1)},\eta_{j_1}^{(i_1)},\cdots,\xi_{j_l}^{(i_l)},\eta_{j_l}^{(i_l)},X)=
\nonumber\\&=&
b_{l-1}(\xi_{j_1}^{(i_1)},\eta_{j_1}^{(i_1)}\cdots,\xi_{j_{l-1}}^{(i_{l-1})},
\eta_{j_{l-1}}^{(i_{l-1})}, X+\xi_{j_{l}}^{(i_l)})\nonumber\\
&&-b_{l-1}(\xi_{j_2}^{(i_2)},\eta_{j_2}^{(i_2)},\cdots,
\xi_{j_{l}}^{(i_l)},\eta_{j_{l}}^{(i_l)},X), \quad l>1 \label{Cf}
\end{eqnarray*}
and the initial function $c_1(\xi_1^{(i)},\eta_1^{(i)},X)=\xi_1^{(i)}$.
\end{Pro}

To construct the hierarchy of the Lax equations we need to expand the coefficients of
operator (\ref{SA}) at $X\to\infty$ and truncate at the required degree. When $n \geq
2$, the expansion of $N_l(\xi_1,\eta_1,\xi_2,\eta_2,\ldots, \xi_l, \eta_l;X)^{-1}$ at
$X\to \infty$ is of the form
\begin{eqnarray*}
&&N_l(\xi_1,\eta_1,\xi_2,\eta_2,\ldots, \xi_l, \eta_l;X)^{-1}
\\&=&
\frac{1}{n X^{n-1} (\sum_{i=1}^l \xi_i)} \sum_{j\geq 0}\left\{\frac{\sum_{i=1}^l
\eta_i}{n X^{n-1} (\sum_{i=1}^l \xi_i)}\right.
\\&&-\left.\frac{1}{n}\sum_{k=0}^{n-2}C^k_n \ X^{k+1-n} (\sum_{i=0}^l \xi_i)^{n-k-1}\right\}^j,\quad n\geq 2.
\end{eqnarray*}
Therefore, if we want to prove that the coefficients of operators (\ref{SA}), i.e.
the functions $a_l$,  are quasi-local defined similarly as definition \ref{local}, we
need to show that the functions $c_l$ can be split into the sum of the image of $D_x$
and the image of $D_y$. It is clear
 that $a_1^{(i)}(\xi_1^{(i)},\eta_1^{(i)},X)$ are quasi-local since
we have $c_1(\xi_1^{(i)},\eta_1^{(i)},X)=\xi_1^{(i)}$. When $l>1$, we have
\begin{Pro}\label{Pro3}
The functions
$c_l(\xi_{j_1}^{(i_1)},\eta_{j_1}^{(i_1)},\cdots,\xi_{j_l}^{(i_l)},\eta_{j_l}^{(i_l)},X)$
for $l>1$ vanish after substitution
$$\xi_{j_1}^{(i_1)}=-\xi_{j_2}^{(i_2)}-\cdots -\xi_{j_{l-1}}^{(i_{l-1})}\quad \mbox {and\quad}
\eta_{j_1}^{(i_1)}=-\eta_{j_2}^{(i_2)}-\cdots -\eta_{j_{l-1}}^{(i_{l-1})}. $$
\end{Pro}

In fact, this proposition does not lead to our intended conclusion
that $b_l$ and thus $a_l$ are quasi-local since the objects are
rational, not polynomial. For example, the expression
$u^2(\frac{\eta_2}{\xi_2}-\frac{\eta_1}{\xi_1})$ representing $ u
\Theta u-(\Theta u) u$ satisfies the above proposition. However, we
can not write $u \Theta u-(\Theta u) u=D_x f_1+D_y f_2$, where both
$f_1$ and $f_2 $ are in $\cR_\Theta$.

Notice that the formulae in Proposition \ref{Pro1} are without
symmetrisation. Combining these expressions, we can obtain the
formulae of high degree terms of operator $S$ when dependant
variables are commuting. Every term is such $S$ is quasi-local. This
implies that every term in $S^n$ is quasi-local. From Theorem
\ref{Th1} follows
\begin{The}\label{qua21}
The hierarchies of commutative ($2+1$)-integrable equation with scalar Lax operators
are quasi-local.
\end{The}

The above setting up is valid for the noncommutative case except Theorem \ref{qua21}.
However, the extension of the concept of quasi-locality to the noncommutative case is
rather complicated. $D_x$ and $D_y$ are the only derivations for the commutative
differential ring. The extension simply enables us to apply $D_x^{-1}$ and $D_y^{-1}$
on the derivations. We know that the commutators are also derivations for a
noncommutative associative algebra and we need to take them into consideration. There
are some further discussions on this topic in \cite{wang21}.

\section*{Summary and discussion}

In this article we have reviewed some recent developments  in the symmetry
approach in the symbolic representation. In particular we have discussed two
different methods. One method is based on the study of conditions for the existence
of a formal recursion operator, another one is based on the explicit analysis of  approximate
symmetries of low degrees.

The formal recursion method is a very efficient tool for testing  the
integrability of a given PDE. This method is based only on the fact of existence of
an infinite hierarchy of higher symmetries and is not sensitive to possible lacunas in
the hierarchy of symmetries. It is extendable to wide classes of equations, including
certain types of non-local equations. The method is rather simple and convenient for
classification of integrable systems of a fixed order. We have illustrated its power
in applications to classification of integrable generalisations of Boussinesq,
Benjamin-Ono and Camassa-Holm type equations.

Explicit analysis of approximate symmetries relies on the structure of dispersion
relations (linear terms) of systems.  We have seen that the structure of symmetries
of a given equation is parametrised by their dispersion relations and the analysis of
existence of approximate symmetries is based on divisibility properties of special
polynomials determined by the dispersion laws. Such divisibility properties are often
obtained via algebraic geometry and number theoretic methods. It gives a detailed
information on the structure of the hierarchy of higher symmetries. So far it  is the
only method which prove to be suitable for a global classification. Here we mean a
classification of integrable equations in all orders. In the frame of this method it
has been demonstrated that any integrable homogeneous evolutionary equation
\[u_t=u_n+f(u_{n-1},\ldots ,u) \qquad n\ge 2,\]
where $w(u)\geq 0$ is a symmetry (a member of a hierarchy) of one of the equation of
order $2,3$ or $5$ presented in Theorem \ref{globclass}.

Systems of equations is a considerably more complicated object. Here the only
``global'' result available is a classification of integrable homogeneous Boussinesq
type equations of odd order (Section \ref{oddorderbousseq}). In the theory of
integrable systems the description of admissible structures of linear terms (the
dispersion laws) for equations and their symmetries is important and yet unsolved
problem. It is so called {\em spectrum} or {\em dispersion} problem. Let us consider
system of PDEs
\begin{equation}\label{sysNeq}
  \bu_t =A\bu_n+\bF(\bu_{n-1},\ldots ,\bu),\quad \bu=(u_1,\ldots,u_N)^T , \quad n\ge 2
\end{equation}
where $A$ is a constant $N\times N$ matrix. Its higher symmetry, if it exists, is of
the form
\[ \bu_{\tau_m} =B(m)\bu_m+\bG(\bu_{m-1},\ldots ,\bu) , \qquad m\ge 2\]
where $B(m)$ is a constant matrix commuting with $A$. Let us denote $\lambda_1,\ldots
,\lambda_N$ the eigenvalues of the matrix $A$ and assume that $\lambda_1\ne0$, then
the set $S_A=\{\lambda_2/\lambda_1, \ldots , \lambda_N/\lambda_1\}$ is called the
{\em spectrum} of system (\ref{sysNeq}). Similarly defined
$S_{B(m)}=\{\mu_2(m)/\mu_1(m), \ldots , \mu_N(m)/\mu_1(m)\}$ the spectrum of
symmetry, where $\mu_1(m),\ldots ,\mu_N(m)$ are eigenvalues of the matrix $B(m)$. The
spectrum is invariant with respect to any re-scaling. In many respects it reflects
properties  of symmetries, conservation laws and solutions of the system. For
example, the existence of higher conservation laws for a system of two even order
equations implies that $S_A=\{-1\}$  (see  \cite{mr87h:35313}).

Integrable system \cite{mr84m:35104}
\[
\left\{\begin{array}{l}
u_t=(5-3\s5)u_3-2uu_1+(3-\s5)vu_1+2uv_1+(1+\s5)vv_1,\\
v_t=(5+3\s5)v_3+(1-\s5)uu_1+2vu_1+(3+\s5)uv_1-2vv_1,\\
\end{array}\right.
\]
has spectrum $S_A=\{\lambda_2/\lambda_1\}=\{-\frac{1}{2}(7+3\s5)\}$. It possesses an
infinite dimensional algebra of symmetries of orders $m\equiv1,3,7,9\mod10$. The
ratio of parameters $\mu_2(m)/\mu_1(m)$ is given by
\[
\frac{\mu_2(m)}{\mu_1(m)}=\frac{\left(1+\exp\left({\frac{2\pi i}{5}}\right)\right)^m}
{1+\exp\left({\frac{2m\pi i }{5}}\right)}.
\]

Symmetry approach in symbolic representation can be applied to the study of possible
spectra and classification of integrable systems. For systems of two homogeneous
differential polynomial equations of second order the problem has been solved in
\cite{MR2070382}. Recently we have been working on this problem for systems of two
equations of odd order. The results obtained will be published elsewhere. Here we
present two rather non-trivial examples of integrable systems, which we believe are
new.

The following system
{\small\begin{eqnarray*} \left\{\begin{array}{ll}
u_t&=(9-5\S3)u_5+ D_x\left\{2(9-5\S3)uu_2+(-12+7\S3)u_1^2\right\}\\&+2 (3-\S3)u_3v+2
(6-\S3)u_2v_1+2 (3-2\S3)u_1v_2\\&-6(1+\S3)uv_3+
D_x\left\{2(33+19\S3)vv_2+(21+12\S3)v_1^2\right\}
\\&+\frac{4}{5}(-12+7\S3)u^2u_1
+\frac{8}{5}(3-2\S3)(vuu_1+u^2v_1)\\&+\frac{4}{5}(24+13\S3)v^2u_1+\frac{8}{5}
(36+20\S3)uvv_1-\frac{8}{5}(45+26\S3)v^2v_1,
\\ \\
v_t&=(9+5\S3)v_5+
D_x\left\{2(33-19\S3)uu_2+(21-12\S3)u_1^2\right\}\\&-6(1-\S3)u_3v+2(3+2\S3)u_2v_1+2
(6+\S3)u_1v_2
\\&+2(3+\S3)uv_3+
D_x\left\{2(9+5\S3)vv_2-(12+7\S3)v_1^2\right\}\\&-\frac{8}{5}(45-26\S3)u^2u_1
+\frac{8}{5}(36-20\S3)vuu_1+\frac{4}{5}(24-13\S3)u^2v_1
\\&+\frac{8}{5}(3+2\S3)(v^2u_1+uvv_1)-\frac{4}{5}(12+7\S3)v^2v_1
\end{array}\right.
\end{eqnarray*}}
possesses an infinite dimensional algebra of higher symmetries
with
 \[\frac{\mu_2(m)}{\mu_1(m)}=\frac{(1+\exp({\frac{\pi i}{6}}))^m}
 {1+\exp({\frac{m\pi i}{6}})}, \qquad m\equiv1,5,7,11\mod\,\, 12.\]

System
\begin{eqnarray*}
\left\{\begin{array}{ll}
u_t&=-\frac{5}{3}u_5-10vv_3-15v_1v_2+10uu_3+25u_1u_2-6v^2v_1\\&+6v^2u_1+12uvv_1
-12u^2u_1,\\
v_t&=15v_5+30v_1v_2-30v_3u-45v_2u_1-35v_1u_2-10vu_3\\&-6v^2v_1+6v^2u_1+12u^2v_1

+12vuu_1.
\end{array}\right.\end{eqnarray*}
possesses symmetries of orders $m\equiv1, 5 \mod\ 6$ with
\[
\frac{\mu_2(m)}{\mu_1(m)}=\frac{(1+\exp({\frac{\pi i}{3}}))^m}
{1+\exp({\frac{m\pi i}{3}})}.\]
There is a reduction $v=0$ to the Kaup-Kupershmidt equation.

\section*{ Acknowledgments}

We would like to  thank J. A. Sanders for useful discussions. AVM also thanks the RFBR for a partial support (grant  05-01-00189). VSN was funded by the EPSRC Postdoctoral Fellowship grant.


\end{document}